\documentclass{aa}

\usepackage[normalem]{ulem}
\usepackage{graphicx}
\usepackage{txfonts}
\usepackage{lipsum}
\usepackage{subcaption}         
\usepackage{xcolor}             
\usepackage{lscape}             
\usepackage{placeins}           

\usepackage{natbib}       
\usepackage[colorlinks=true, allcolors=blue]{hyperref}  

\begin{document}

   \title{Tracing Lyman alpha escape in the CRISTAL-02 galaxy at z$\sim$5.3}
   \titlerunning{CRISTAL-02 spatial study}


\author{
  Arshi Ali\inst{1,2}\thanks{\email{a.ali@uandresbello.edu}},
  Lucia Guaita\inst{1,2},
  Manuel Aravena\inst{3,2},
  Rebecca L. Davies \inst{4},
  Jorge González \inst{5,2},
  Vicente Villanueva\inst{3},
  Diego Oyarzún \inst{1},
  Rodrigo Herrera-Camus \inst{6,2},
  Ambra Nanni \inst{7,8},
  Andreas Faisst \inst{9},
  Anton M. Koekemoer\inst{10},
  Deanne B. Fisher \inst{4},
  Hanae Inami \inst{11},
  Juan Molina \inst{12,2},
  Justin Spilker\inst{13}, 
  Lun-Jun Liu \inst{9},
  Manuel Solimano \inst{14},
  Michele Ginolfi \inst{15,16},
  Michael Romano \inst{17,18},
  Negin Nezhad \inst{19},
  Poulomi Dam \inst{20}, 
  Wuji Wang\inst{9},
  Yuan Li \inst{13}
}
\authorrunning{Ali et al. 2026}

\institute{
  Universidad Andres Bello, Facultad de Ciencias Exactas, 
  Departamento de Fisica y Astronomia, Instituto de Astrofisica, 
  Fernandez Concha 700, Las Condes, Santiago RM, Chile
  \and
  Millennium Nucleus for Galaxies (MINGAL)
  \and
  Instituto de Estudios Astrof\'{\i}cos, Facultad de Ingenier\'{\i}a y Ciencias, Universidad Diego Portales, Av. Ej\'ercito 441, Santiago, Chile
  \and
  Centre for Astrophysics and Supercomputing, Swinburne University of Technology, John Street, Hawthorn, 3122 VIC, Australia
  \and
  Instituto de Astrofísica, Facultad de Física, Pontificia Universidad Católica de Chile, Santiago 7820436, Chile
  \and 
  Universidad de Concepci\'on, Barrio Universitario, Concepci\'on, Chile
  \and
  National Centre for Nuclear Research, ul. Pasteura 7, 02-093 Warsaw, Poland 
  \and
  INAF - Osservatorio Astronomico d’Abruzzo, Via Maggini SNC, 64100 Teramo, Italy
  \and
  Caltech/IPAC, 1200 E. California Blvd. Pasadena, CA 91125, USA
  \and 
  Space Telescope Science Institute, 3700 San Martin Drive, Baltimore, MD 21218, USA
  \and
  Hiroshima Astrophysical Science Center, Hiroshima University, 1-3-1 Kagamiyama, Higashi-Hiroshima, Hiroshima 739-8526, Japan
  \and
  Instituto de F\'{i}sica y Astronom\'{i}a, Universidad de Valpara\'{i}so, Avda. Gran Breta\~{n}a 1111, Valpara\'{i}so, Chile
  \and
  Department of Physics and Astronomy and George P. and Cynthia Woods Mitchell Institute for Fundamental Physics and Astronomy, Texas A\&M University, 4242 TAMU, College Station, TX 77843-4242, US
  \and
  Centro de Astrobiolog\'ia (CAB), CSIC-INTA, Ctra. de Ajalvir km 4, Torrej\'on de Ardoz, E-28850, Madrid, Spain.
  \and
  Dipartimento di Fisica e Astronomia, Università degli Studi di Firenze, Via G. Sansone 1,I-50019, Sesto Fiorentino, Firenze, Italy
  \and
  INAF - Osservatorio Astrofisico di Arcetri, Largo E. Fermi 5, I-50125, Firenze, Italy
  \and
  Max-Planck-Institut für Radioastronomie, Auf dem Hügel 69, 53121 Bonn, Germany
  \and
   INAF - Osservatorio Astronomico di Padova, Vicolo dell’Osservatorio 5, I-35122 Padova, Italy
  \and
  Department of Physics and Astronomy, University of California, Riverside, 900 University Avenue, Riverside, CA 92521, USA
  \and
  Dipartimento di Fisica e Astronomia Galileo Galilei, Universit{\`a} degli Studi di Padova, Vicolo dell’Osservatorio 3, 35122 Padova, Italy
  }


 
\abstract
{
\textit{Aim}. We investigate the mechanisms regulating the escape of  Lyman-alpha (Ly$\rm \alpha$) photons in the star-forming galaxy CRISTAL-02 at redshift z $\approx$ 5.3. Previous studies revealed a diffused and extended Ly$\rm \alpha$ emitting gas (radius$\sim$16.5 kpc) and a possible [C\,\textsc{ii}] outflow. The galaxy has been identified with five distinct clumps, two of which have been suggested as potential active galactic nuclei (AGN) candidates (Clump A and B hereafter). We study how AGN or star formation activities influence the spatial distribution and visibility of Ly$\rm \alpha$ emission.

\textit{Methods}. Using integral field spectroscopic Ly$\alpha$ data from VLT/MUSE and JWST/NIRSpec integral field unit (IFU) observations of H$\alpha$ and H$\beta$, complemented by NIRCam UV imaging, we constructed spatially matched emission-line maps. We derived flux, emission-line ratio, and extinction maps as well as spatially resolved measurements of the Ly$\alpha$ escape fraction and ionizing photon production efficiency.

\textit{Results}. We find that the Ly$\alpha$ emission extends preferentially along the [C\,\textsc{ii}] outflow direction, while the H$\alpha$ and UV emission trace the galactic disk. Two compact regions (Clumps A and B) exhibit contrasting properties: Clump A, associated with lower dust attenuation, shows enhanced Ly$\alpha$/H$\alpha$ ratios and a higher Ly$\alpha$ escape fraction, whereas Clump B shows stronger H$\alpha$ and UV emission but suppressed Ly$\alpha$. The results indicate that Ly$\alpha$ escape can be influenced by AGN or star formation-driven outflows, dust attenuation, and local gas conditions.

\textit{Conclusions}. This study provides one of the first spatially resolved measurements of the Ly$\alpha$ escape fraction and ionizing photon production efficiency in distant systems, and it highlights the role of feedback-driven gas clearing and anisotropic outflows in shaping the observed Ly$\alpha$ properties. The current observational evidence is most consistent with the presence of an outflow, while the mechanism driving it remains uncertain, as the available data do not uniquely distinguish between AGN and star formation-driven feedback. Future spatially resolved multiwavelength IFU observations combining Ly$\alpha$, H$\alpha$, and [C\,\textsc{ii}] emission will be essential to disentangling these scenarios and establishing the dominant feedback mechanism.
}


\maketitle

\section{Introduction}

The Ly$\rm \alpha$ emission line is one of the most powerful tools for studying early galaxies. \citep[e.g.,][]{Rhoads2000, Kashikawa2011, Matthee2015, Zheng2017, Taylor2020, Witstok2024}. Lyman-alpha  photons are produced when ionized hydrogen recombines and subsequently undergoes electronic transitions in neutral hydrogen (HI). As a result, Ly$\rm \alpha$  emission is closely linked to star-forming regions around young massive stars, while they are strongly shaped by the surrounding H\,\textsc{i} through resonant scattering. Because Ly$\rm \alpha$ is a resonant transition, photons undergo multiple scatterings with H\,\textsc{i} before escaping, leading to complex spatial morphologies that often extend well beyond the stellar component \citep[e.g.,][]{Wisotzki2016, Leclercq2017}. Understanding the mechanisms that regulate Ly$\rm \alpha$ escape is therefore crucial for interpreting high-redshift galaxy surveys and for constraining models of galaxy formation and evolution \citep{Dijkstra2014, Inoue2018}.

Despite its diagnostic potential, interpreting Ly$\rm \alpha$ emission is nontrivial. The resonant scattering process makes the line highly sensitive to gas geometry, velocity fields, and viewing angle, while each scattering event increases the likelihood of absorption by interstellar dust grains. These effects often reshape the intrinsic line profile and can impact the overall escape fraction \citep{Gazagnes2020}. In contrast, H$\rm \alpha$ is a non-resonant recombination line that escapes more directly from \ion{H}{II} regions and thus serves as a more reliable tracer of the instantaneous star formation rate and kinematics \citep{Kennicutt1998, Erb2006, Forster2009,  Hayes2010, Clarke2024, Covelo2025, Ren2025}. Comparing Ly$\rm \alpha$ and H$\rm \alpha$ luminosities enables the measurement of the Ly$\rm \alpha$ escape fraction and provides valuable insight into the effects of radiative transfer \citep[e.g.,][]{Mas-Ribas2017}, ionization conditions, and dust attenuation since Ly$\rm \alpha$, being at a shorter wavelength and resonantly scattered, experiences effectively higher dust attenuation than the non-resonant H$\rm \alpha$ line. Furthermore, we can understand the nature of the Ly$\rm \alpha$ emission by complementing it with non-resonant tracers. The [C\,\textsc{ii}] at 158 $\mu$m line provides a powerful alternative, as it is not affected by resonant scattering and thus offers a more reliable probe of the systemic redshift \citep[e.g.,][]{Capak2015, Carniani2018}. In addition, [C\,\textsc{ii}] emission is commonly used as a tracer of star formation activity \citep[e.g.,][]{Stacey2010, Carilli2013}, enabling a more robust characterization of the host galaxy properties.

The escape of Ly$\rm \alpha$ photons can be strongly modulated by the large-scale gas dynamics and the influence of AGNs. Furthermore, galactic outflows can create low-opacity channels that enhance Ly$\rm \alpha$ escape, whereas the denser disk regions may trap photons and reduce the observed flux. Feedback driven by AGNs feedback further shapes the interstellar medium (ISM) by ionizing gas and driving energetic winds that redistribute dust and gas, potentially creating pathways for Ly$\rm \alpha$ photons to escape more efficiently \citep{Duval2016, Liu2022}. The geometry of the gaseous disk and the orientation of these outflows relative to the observer can lead to anisotropic Ly$\rm \alpha$ emission and complex line profiles \citep{Verhamme2015, Verhamme2018}. Spatially resolved observations are therefore essential to disentangling these effects and quantifying the relative contributions of star formation and AGN activity to Ly$\rm \alpha$ visibility.

In this paper, we present a joint spatially resolved analysis of the Ly$\rm \alpha$, H$\rm \alpha$, and [C\,\textsc{ii}] emission in the CRISTAL-02 galaxy at $z\approx5.3$ based on recent James Webb Space Telescope (JWST), Multi Unit Spectroscopic Explorer (MUSE) and Atacama Large Millimeter/submillimeter Array (ALMA) observations. We investigate the resolved distribution of the Ly$\rm \alpha$ escape fraction and ionizing photon production efficiency in the context of the nebular extinction and gas morphology, aiming to connect Ly$\rm \alpha$ radiative transfer with the physical conditions of the ISM.
The paper is structured as follows: In Section~\ref{Data}, we introduce the CRISTAL-02 system, describing previous studies focusing on this source; the datasets used in this work; and the methodology adopted to construct the Ly$\alpha$, H$\alpha$, H$\beta$, and UV emission line maps. Section~\ref{results} presents the results of our spatially resolved analysis, including the morphology and emission-line maps of the galaxy. In Section~\ref{discussion}, we discuss the implications of our results, focusing on the Ly$\alpha$ escape fraction, ionization efficiency, and the nature of the Ly$\alpha$ emission. Finally, our main conclusions are summarized in Section~\ref{conclusion}.

Throughout this paper, We adopt a flat Lambda Cold Dark Matter ($\Lambda$CDM) cosmology with $\rm H_0 = 70~\mathrm{km\,s^{-1}\,Mpc^{-1}}$, $\rm\Omega_{\mathrm{m}} = 0.3$, and $\rm\Omega_{\Lambda} = 0.7$. We use the systemic redshift derived from the [C\,\textsc{ii}] observations presented in the CRISTAL survey papers, adopting $\rm z_{\mathrm{[C\,II]}} = 5.293$. At this redshift, $1^{\prime\prime}$ corresponds to $\sim$~6.1~kpc.

\section{{Data and methodology}}
\label{Data}
\subsection{Previous studies of the source}
\label{previous_study}

The target CRISTAL-02 (also cataloged as DC-848185 in \citealt{Laigle2016}, LBG-1 in \citealt{Riechers2014}, and HZ6 in \citealt{Capak2015}) lies at $z\sim5.3$ in the overdense environment surrounding the AzTEC–3 submillimeter galaxy (SMG) in the Cosmic Evolution Survey Deep Field (COSMOS; \citealt{koekemoer2007, Scoville2007, Capak2011, Riechers2014}). Early MUSE tomography of the AzTEC–3 region revealed a compact overdensity of Ly$\alpha$ sources and extended Ly$\alpha$ emission connected to interactions in the inner core of the protocluster, suggesting ongoing gas-rich interactions and extended reservoirs of H\,\textsc{i} around the SMG and its companions \citep{Guaita2022}. These MUSE results motivate a spatially resolved comparison between resonant Ly$\alpha$ and non-resonant recombination tracers to map the neutral and ionized gas topology in a protocluster environment.

The ALMA-CRISTAL survey targets $4 \lesssim z \lesssim 6$ main-sequence and UV-selected galaxies selected from the ALPINE sample \citep{Capak2015, LeFevre2020, Faisst2020a}. The survey provides high-resolution [C\,\textsc{ii}] and dust continuum imaging, enabling resolved maps of cold gas and obscured star formation on $\sim$kiloparsec (kpc) scales \citep{HerreraCamus2025}. Across the sample, extended [C\,\textsc{ii}] structures are observed, often associated with tidal or interaction features as well as evidence of fast, spatially offset [C\,\textsc{ii}] wings or outflows in several systems (\citealt{HerreraCamus2021, HerreraCamus2025, Birkin2025, Davies2026}). These results provide an essential cold-gas reference frame for comparison with ionized gas tracers and for identifying kinematic axes (disk versus outflow).

CRISTAL-02, one of the targets in the survey, shows a complex [C\,\textsc{ii}] morphology and kinematics in high-resolution ($\sim 0.2\arcsec$) ALMA imaging \citep{HerreraCamus2025, Lee2025}. Its extended [C\,\textsc{ii}] emission is consistent with the broader trends seen in the sample while also revealing system-specific features that are particularly relevant for studying the connection between cold gas, outflows, and Ly$\alpha$ emission.

Recently, the ALPINE-CRISTAL-JWST program \citep{Faisst2026} identified multiple broad H$\alpha$ emitters and type-1 AGN candidates among the $z\sim4{-}6$ sample, with at least two such candidates associated with CRISTAL-02 \citep{Ren2025}. In this work, we adopted the clump nomenclature used by \citet{Ren2025} and refer to these regions as Clump A and Clump B. The nomenclature differs in \citet{Davies2026}, where our Clump A and Clump B are referred to as Clump C and Clump A, respectively. The evidence of AGN activity in this system remains inconclusive. In contrast, CRISTAL-02 exhibits a prominent and well-supported outflow \citep{Davies2026}, making it one of the most interesting cases within the CRISTAL sample. Such an outflow provides a natural mechanism for shaping the gas geometry, regulating the dust distribution, and creating low column density channels that may facilitate anisotropic Ly$\alpha$ escape.

It is important to note that given the spatial resolution of our MUSE-matched data, the individual clumps in the southern region (Clumps B, C, and D in the nomenclature of \citealt{Ren2025}) are not always clearly distinguishable. In particular, the region we refer to as Clump B may contain blended emission from multiple physically distinct components, including the potential AGN candidate and the region associated with the outflow. In this context, CRISTAL-02 represents a particularly valuable system for investigating how Ly$\alpha$ radiative transfer is influenced by feedback-driven gas structure. While a potential contribution from AGNs cannot be fully excluded, evidence of AGN activity remains indirect and is primarily based on the detection of broad H$\alpha$ components reported by \cite{Ren2025}. However, the spatially resolved line ratios and Ly$\alpha$ morphology presented here can be consistently explained within a framework where stellar feedback and outflows dominate. If AGN activity is confirmed, it would likely act to further enhance these feedback processes and facilitate Ly$\alpha$ escape.

\subsection{Ly$\rm \alpha$ emission from MUSE IFU}

To construct the Ly$\alpha$ emission map, we used the archival VLT/MUSE data cube presented in \cite{Guaita2022}, who previously analyzed the Ly$\alpha$ emission in this galaxy (we refer to that paper for a detailed description of the data acquisition and the reduction procedures). The MUSE data have a pixel scale of 0.2\arcsec\ per spaxel and a point spread function (PSF) with full width half maxima (FWHM) $\approx 0.74$\arcsec\ at the observed Ly$\alpha$ wavelength. The Ly$\alpha$ narrow-band image was produced by stacking the cube over the spectral range 7655--7670\AA, following the wavelength interval adopted by \cite{Guaita2022} to capture the bulk of the high signal-to-noise (S/N) Ly$\alpha$ emission associated with the main Ly$\alpha$ peak. This spectral window lies redward of the systemic Ly$\alpha$ wavelength ($\sim7650.2$\AA\ at $z=5.293$), reflecting the observed redshifted Ly$\alpha$ emission.

Prior to this step, a first-order (linear) continuum model was fitted and subtracted spaxel-by-spaxel to obtain a continuum subtracted cube. Then, spectral channels within this wavelength interval were combined using inverse-variance weighting derived from the MUSE variance cube such that channels with a lower noise were given a higher weight, ensuring that more reliable data dominate the final Ly$\alpha$ flux map. The selected spectral window corresponds to a velocity interval of approximately 588 km s$^{-1}$ centered on the Ly$\alpha$ line ($\sim$7650.2 \AA). The resulting narrowband map has a $1\sigma$ rms sensitivity of $\sim0.18\times10^{-20}~$erg~s$^{-1}$~cm$^{-2}$, as measured from the line-free region of the MUSE image.

\subsection{H$\rm \alpha$ and H$\rm \beta$ emission from JWST/NIRSpec IFU}

CRISTAL-02 was observed with JWST/Near Infrared Spectrograph (NIRSpec) IFU as part of a Cycle 2 GO program targeting 18 star-forming galaxies at z$\simeq4-6$ \citep[Program ID: 3045,][]{ Faisst2026}. The observations were carried out using the G235M/F170LP (1.7–3.1 $\mu$m, R$\sim$1000)
and G395M/F290LP (2.9–5.1 $\mu$m, R$\sim$1000) dispersers. Details of the observations, data reduction, calibration, and processing procedures are presented in \citet{Fujimoto2025} and \citet{Faisst2026}. In this work, we make use of the G395M reduced data cube, which covers the key rest-frame optical emission lines, including H$\rm \alpha$ and H$\rm\beta$, at the redshift of CRISTAL-02 required for the analysis of this work.

Since the spatial resolution of JWST ($\mathrm{FWHM} \approx 0.1''$) differs significantly from that of MUSE ($\mathrm{FWHM} \approx 0.7''$), the JWST data cube was convolved to match the MUSE PSF. Assuming both PSFs can be approximated as Gaussians, we constructed a two-dimensional Gaussian kernel with a width given by $\sigma_{\rm kernel} = \sqrt{\sigma_{\rm MUSE}^2 - \sigma_{\rm JWST}^2}$, where $\sigma = \mathrm{FWHM}/(2\sqrt{2\ln2})$. The kernel width was converted to pixel units using the pixel scale of the JWST data ($\rm\sim0.1''/$per pixel), and the convolution was applied using a fast fourier transform (FFT)-based method implemented in the \texttt{astropy.convolution} package. This procedure ensures that the final JWST cube is matched to the spatial resolution of the MUSE observations while preserving flux. After convolution, we reprojected the JWST cube to align with the MUSE coordinate grid, ensuring a common reference frame.

We performed continuum subtraction on the reprojected NIRSpec IFU cube and modeled the H$\alpha$ and H$\beta$ emission lines on a spaxel-by-spaxel basis. The continuum in each spaxel was estimated by fitting a third-order polynomial to the full spectral range, capturing the global shape of the spectrum while avoiding overfitting. The fit was performed using a least-squares minimization, with the initial normalization set by the median flux level. The resulting best-fitting continuum model was subtracted from the original spectrum to produce a continuum-subtracted data cube, which was used for subsequent emission-line analysis.

For the H$\alpha$ region, we accounted for blending with the adjacent [N\,\textsc{ii}] $\lambda\lambda$6549, 6585 lines by performing a simultaneous multicomponent Gaussian fit within a fixed spectral window of $\rm\pm 0.02~\mu$m around the expected observed wavelength of H$\alpha$. The emission lines were modeled using single Gaussian components, with the centroids and widths of H$\alpha$ and [N\,\textsc{ii}] tied to share the same kinematic properties. The wavelength separations between the lines were fixed to their laboratory values, and the flux ratio of [N\,\textsc{ii}] $\lambda$6549 to $\lambda$6585 was fixed to the theoretical value of 1:3 \citep{Acker1989}. Initial parameter guesses were set using the peak flux in the fitting window, with the [N\,\textsc{ii}] amplitude initialized to 50$\%$ of the H$\alpha$ amplitude. The fitting was performed using a Levenberg–Marquardt least-squares minimization, as implemented in the \texttt{astropy.modeling} framework. Line fluxes were computed analytically from the best-fitting Gaussian parameters as $F = A \sigma \sqrt{2\pi}$, where $A$ and $\sigma$ are the amplitude and standard deviation of the Gaussian profile. For the H$\beta$ line, a single Gaussian component was fitted independently within a similar spectral window after continuum subtraction, with all parameters treated as free.

We note that although broad H$\rm \alpha$ emission has been reported for this source based on higher spatial-resolution NIRSpec IFU observations \citealt{Davies2026, Ren2025}), the S/N of the MUSE convolved image does not allow for a robust multicomponent decomposition. We therefore adopted single-Gaussian fits to characterize the total H$\rm \alpha$ and H$\rm \beta$ fluxes.

\subsection{UV emission from JWST/NIRCam}

In this work, we utilized the reduced and calibrated NIRCam imaging products observed as a part of the COSMOS-Web survey \citep[Program ID: 1727,][]{Casey2023} in the F115W, F150W, F277W, and F444W filters. Details of the observations and data reduction are described in \cite{HerreraCamus2025}.

To study the rest-frame UV emission of the galaxy, we specifically used F115W filter image from NIRCam. We initially tried combining the HST (F105W, F125W, F160W) and JWST/NIRCam imaging (F115W, F150W) to construct a rest-frame UV continuum and UV spectral slope ($\rm\beta$) map, assuming a power-law UV spectrum. However, the lower S/N of the HST data introduced significant uncertainties, particularly for pixel-by-pixel UV slope measurements. After PSF matching, the resulting $\rm\beta$ map was found to be unreliable. We therefore adopted a conservative approach and constructed the UV map using only the JWST/NIRCam F115W filter, which probes the rest-frame far-UV emission ($\rm\lambda_{rest}\approx1800\AA$) at the redshift of the galaxy. The F115W image was thus used as a proxy for the UV continuum emission. Following the similar approach used for the NIRSpec IFU data cube, the NIRCam F115W image was first convolved and reprojected to match the spatial scale of the MUSE cube. This allowed us to generate UV emission maps that are directly comparable to the MUSE emission line maps, facilitating spatial analyses.

\begin{figure*}
\centering
\includegraphics[width=16cm, trim= 0 0cm 0 0]{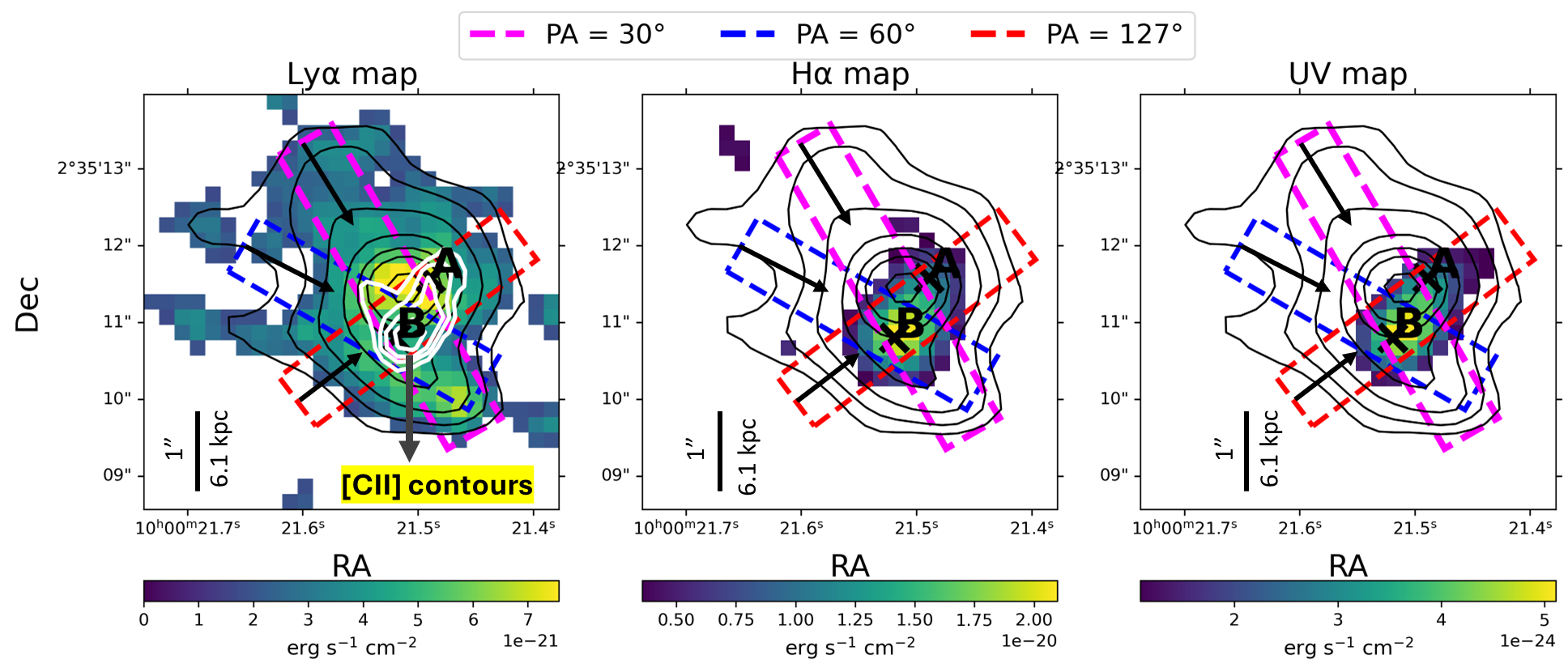}
\includegraphics[width=16cm, trim= 0 0cm 0 0]{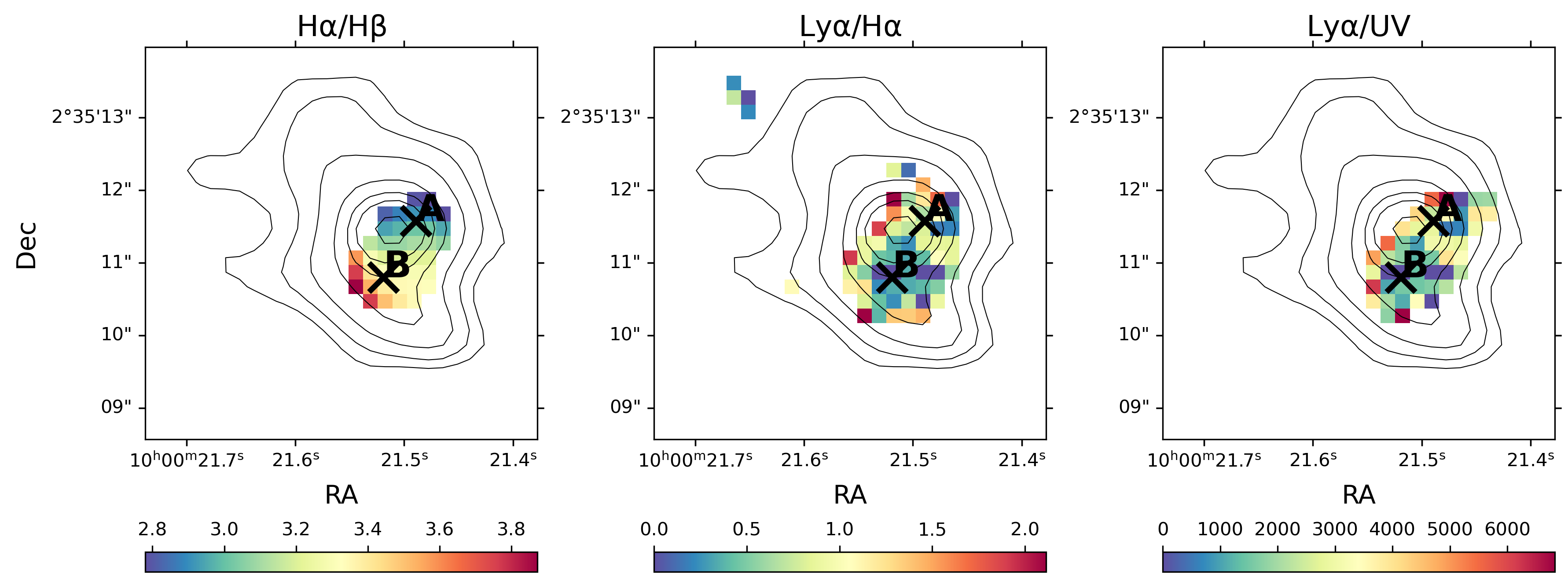}
\caption{\small Top row: Ly$\rm \alpha$, H$\rm \alpha$, and UV flux maps. The Ly$\rm \alpha$ image is smoothed with a Gaussian kernel of $\sigma$ = 1.5 pixel for display purposes, while all quantitative analysis is performed on the unsmoothed data. For H$\rm \alpha$, and UV flux maps, a S/N $>$ 2 threshold is applied to ensure reliable emission detection. The black contours correspond to the 60th, 75th, 85th, 92nd, 96th, 98th, and 99.6th percentiles of the smoothed Ly$\alpha$ flux distribution. The white contours on Ly$\rm \alpha$ map represent the Briggs-weighted [C\,\textsc{ii}] moment-0 map with levels $[4, 7, 10, 15]\times\sigma$, where $\sigma = 0.025$ Jy beam$^{-1}$ km s$^{-1}$ corresponds to the rms noise of the [C\,\textsc{ii}] map. The red dashed strip marks the orientation of the galactic disk (PA=127$^\circ$), the blue dashed strip indicates the direction of the [C\,\textsc{ii}] outflow (PA=60$^\circ$) as described in \cite{Davies2026}, and the magenta dashed strip at PA=30$^\circ$ captures the opening angle of the outflow. The two black crosses denote the positions of the AGN candidates reported by \cite{Ren2025}. The black arrows mark the slits orientation.
Bottom row: H$\rm \alpha$/H$\rm \beta$, Ly$\rm \alpha$/H$\rm \alpha$ and Ly$\rm \alpha$/UV line ratio maps with Ly$\rm \alpha$ contours over-plotted on the top .
}
\label{lyaHa_fluxmap}
\end{figure*}

\section{Results}
\label{results}

\subsection{Morphology}

In Figure~\ref{lyaHa_fluxmap} (top), we present the Ly$\rm \alpha$ map with black contours tracing Ly$\rm \alpha$ emission and white contours tracing [C\,\textsc{ii}] emission (refer to \citealt{Davies2026}, for more details). Alongside, the PSF-matched H$\rm \alpha$ map and UV map are shown with Ly$\rm \alpha$ contours, illustrating the spatial correspondence between the emission lines. The  Ly$\rm \alpha$ emission map was smoothed using a Gaussian kernel with a sigma of 1.5 pixels in order to improve the S/N of extended structures while preserving the overall morphology. This smoothing was applied solely for visualization purposes and does not affect any quantitative measurements, which are based on the original unsmoothed data. A mask for the Ly$\alpha$ map was defined using pixels above the $\sim$60th percentile of the smoothed flux distribution, corresponding approximately to a S/N threshold that selects significant emission while excluding noise-dominated regions. For H$\rm \alpha$ and UV maps, we applied a mask with $\mathrm{S/N} > 2$ to ensure the robustness of the detected emission. The two black crosses in all the panels in Figure~\ref{lyaHa_fluxmap} mark the locations of the possible AGN candidates identified by \cite{Ren2025}.

The slits overlaid in the figure correspond to the orientations of the [C\,\textsc{ii}] outflow (PA = 60°) and the disk (PA = 127°), as defined by \cite{Davies2026}. We added an additional slit to better capture the geometry of the Ly$\alpha$ emission relative to the [C\,\textsc{ii}] outflow. While the PA $= 60^\circ$ slit traces the direction of the outflow, the PA $= 30^\circ$ slit accounts for its opening angle when considering the bicone geometry described in \cite{Davies2026}. The measured half-opening angle of the biconical structure is 15–30$^\circ$. This orientation therefore plausibly traces the outer edge of the outflow cone. The peak of the Ly$\alpha$ emission visible toward the south is aligned with both the PA $= 60^\circ$ and PA $= 30^\circ$ slits.

A comparison of the emission-line morphologies showed that Ly$\alpha$ is significantly more extended than the H$\alpha$ and UV emission. The H$\alpha$ emission extends to 13.8~kpc toward the northwest, consistent with the UV morphology, which reaches $\sim$14~kpc in the same direction. In contrast, the Ly$\alpha$ exhibits a diffuse and extended structure, reaching up to $\sim$33~kpc toward the northeast and aligning with the [C\,\textsc{ii}] outflow. The geometry of the Ly$\alpha$ and [C\,\textsc{ii}] emission is discussed in more detail in Section~\ref{lya_nature}.

The Ly$\alpha$ emission is notably brighter around Clump A, which may indicate more favorable Ly$\alpha$ escape conditions and/or radiative transfer effects in the local ISM, consistent with the higher Ly$\alpha$ escape fractions and Ly$\alpha$/H$\alpha$ ratios measured in this region (see Section~\ref{esc_fraction}). Both the H$\alpha$ and UV maps exhibit emission that extends along the orientation of the galactic disk. Interestingly, Clump~B appears significantly brighter in both tracers compared to the Ly$\alpha$ emission, highlighting this region as a site of recent star formation dominated by young stellar populations, consistent with the stellar age maps that identify it as the youngest region of the galaxy \citep{Davies2026}.

Notably, the peak of the H$\alpha$ and UV emission coincides spatially with the dust continuum distribution \citep[see Figure 11 of][]{HerreraCamus2025}. Under typical conditions, such a configuration would be expected to produce strong attenuation of both UV and nebular emission. However, the observed enhancement of H$\alpha$ and UV emission suggests that dust extinction is not dominant along the line of sight in this region. This may indicate the presence of multiphase mixing between ionized, neutral, and dusty components, possibly driven by turbulent outflow activity that creates low-opacity channels through which radiation can escape. We performed Ly$\alpha$ radiative-transfer modeling using the "redshift Estimator for Line Profiles of Distant Lyman Alpha Emitters" \citep[\texttt{zELDA II};][]{Gurung2022, Gurung2025} (see Appendix~\ref{app:zelda}), which further supports a low-velocity outflow scenario in the direction traced by the [C\,\textsc{ii}] emission.

\subsection{Line ratio maps}

\begin{figure}
\centering
\includegraphics[width=\columnwidth, trim= 0 0cm 0 0]{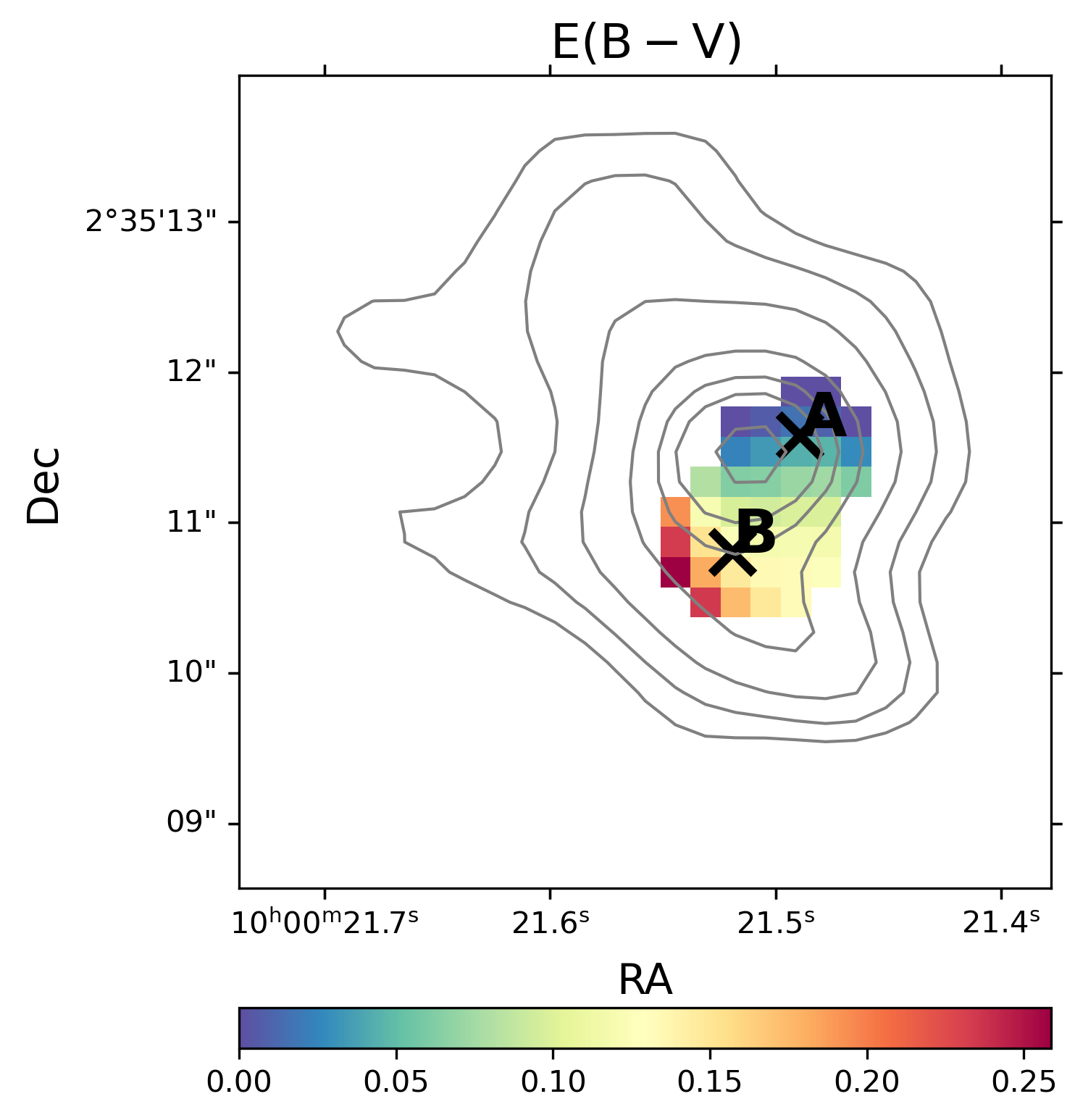}
\caption{\small Spatial distribution of dust attenuation in the galaxy shown through color excess, $\rm E(B-V)$. The positions of the two AGN candidates are marked with black crosses (labeled A and B).}
\label{ebv_Av_maps}
\end{figure}

The Ly$\rm \alpha$/H$\rm \alpha$ and Ly$\rm \alpha$/UV ratios are powerful diagnostics of the Ly$\rm \alpha$ radiative transfer process, as they trace how ionizing photons interact with the surrounding H\,\textsc{i} and dust. In an ideal dust-free case B recombination scenario, the intrinsic Ly$\rm \alpha$/H$\rm \alpha$ ratio is expected to be $\sim$8.7 \citep{Brocklehurst1971, Hummer1987}. Deviations from this value therefore provide direct information on Ly$\rm \alpha$ escape fractions and radiative transfer effects within the interstellar and circumgalactic media. Similarly, the Ly$\rm \alpha$/UV ratio probes the balance between ionizing and nonionizing emission and is often used as an indicator of dust geometry, ISM clumpiness, and feedback-driven outflows \citep[e.g.,][]{Hayes2011, Henry2015, Runnholm2020}. Together, these ratios offer a spatially resolved view of how gas, dust, and radiation are coupled in high-redshift galaxies. Figure~\ref{lyaHa_fluxmap} (bottom) shows the H$\rm \alpha$/H$\rm \beta$, Ly$\rm \alpha$/H$\rm \alpha$ and Ly$\rm \alpha$/UV flux ratio maps.

Both Ly$\alpha$/H$\alpha$ and Ly$\alpha$/UV ratios are systematically higher in the region surrounding Clump A compared to Clump B. With a small region  (1$\arcsec$ aperture radius) centered on the two clumps, we measure median Ly$\alpha$/H$\alpha$ ratios of $0.76 \pm 0.42$ for Clump A and $0.35 \pm 0.25$ for Clump B. Similarly, the Ly$\alpha$/UV ratio is higher in Clump A ($2836 \pm 514$) than in Clump B ($1486 \pm 101$). This indicates that Ly$\rm \alpha$ photons escape more efficiently from the region surrounding Clump A. Such enhanced ratios can result from either a lower H\,\textsc{i} column density or reduced dust extinction, both of which facilitate Ly$\rm \alpha$ escape. In contrast, the lower ratios observed around Clump B suggest a dustier or more optically thick environment, where Ly$\rm \alpha$ photons are more likely to be absorbed or scattered out of the line of sight. Overall, the ratio maps reinforce the morphological trends seen in the Ly$\rm \alpha$ flux distribution, i.e., Clump A likely represents a region of more transparent outflowing gas, whereas Clump B is associated with denser, dustier conditions that inhibit Ly$\rm \alpha$ escape. This is consistent with the dust emission peak seen in \cite{HerreraCamus2025} at the location of Clump B.

Using the H$\rm \alpha$ and H$\rm \beta$ flux maps, we derived the Balmer decrement (H$\rm \alpha$/H$\rm \beta$; Figure~\ref{lyaHa_fluxmap}, bottom left) at the spatial resolution of MUSE on a spaxel-by-spaxel basis and further estimated the nebular color excess, $\rm E(B-V)$ (Figure~\ref{ebv_Av_maps}). For this calculation, we assumed an intrinsic Case B recombination ratio of H$\rm \alpha$/H$\rm \beta$ = 2.86 at $\rm T = 10^4$ K and $\rm n_e = 100$ cm$^{-3}$ \citep{Hummer1987, Osterbrock2006}. The observed ratio was then converted into $\rm E(B-V)$ using the \cite{Calzetti2000} attenuation curve. We found $\rm E(B-V)$ values ranging from approximately 0 to 0.25 mag across the system. Clump A has a relatively low value, $\rm E(B-V)=0.037\pm0.04$, indicating lower dust attenuation, which is consistent with its strong Ly$\alpha$ emission and enhanced photon escape (Section~\ref{esc_fraction}). In contrast, Clump B has a higher value, $\rm E(B-V)=0.142\pm0.05$, suggesting stronger dust attenuation that may preferentially suppress Ly$\alpha$ emission, while the H$\rm \alpha$ and UV emission are comparatively less affected.

The spatial variation observed in the H$\alpha$/H$\beta$ ratio map shows a gradient that may be influenced by the PSF matching applied to the JWST data. In particular, regions A and B show intrinsically different line ratios, and the convolution to the lower MUSE resolution introduces spatial mixing of flux between adjacent spaxels. This effect can produce a smooth transition in the observed ratio, even if the underlying distribution varies more sharply. To verify this change, we compared the H$\alpha$/H$\beta$ map obtained at the native JWST/NIRSpec resolution (see Figure~\ref{hahb_notconvolved}). We find that the contrast between the two regions is enhanced in the native-resolution data, indicating that the gradient is primarily physical, while the convolution acts to smooth the intrinsic variations.

\section{Discussion}
\label{discussion}
\subsection{Escape fraction}
\label{esc_fraction}

\begin{figure*}
\centering
\includegraphics[width=16cm, trim= 0 0cm 0 0]{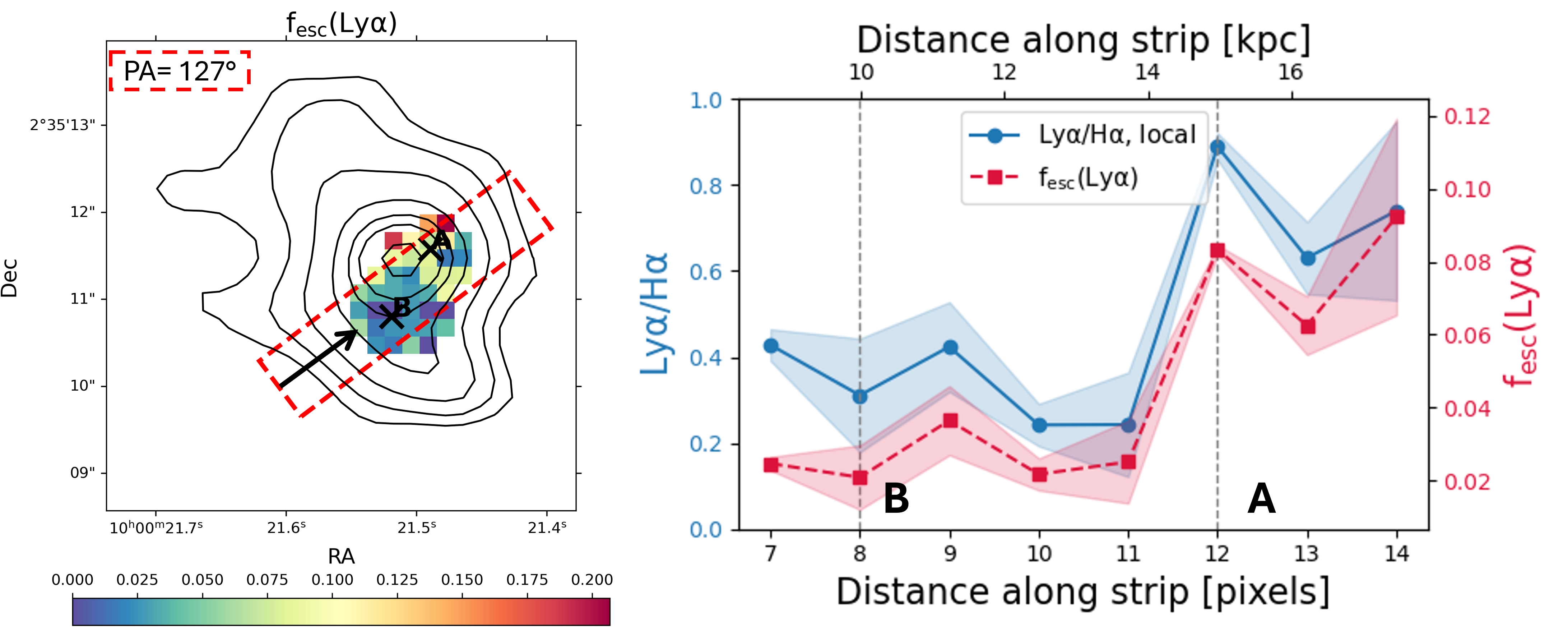}

\caption{\small (Left) Spatial map of the ${\rm Ly\alpha}$ escape fraction ($\rm f_{esc}$) with the red slit indicating the PA= 127$^\circ$ extraction region along the galactic disk. The arrow marks the direction along which the radial profile was derived. The positions of the two AGN candidates (A and B) are highlighted with black crosses. (Right) Radial profiles of the ${\rm Ly\alpha}$/H${\rm \alpha}$ ratio (blue) and the ${\rm Ly\alpha}$ escape fraction ($\rm f_{esc}(Ly\alpha)$; red) measured along the strip. The shaded regions indicate the corresponding  $\rm 1\sigma$ uncertainty. Vertical lines denote the projected locations of the Clump A and B. 
}
\label{f_esc}
\end{figure*}

The Ly$\alpha$ escape fraction ($\rm f_{\rm esc}^{\rm Ly\alpha}$) quantifies the fraction of intrinsically produced Ly$\alpha$ photons that escape a galaxy without being absorbed by dust, despite potentially undergoing multiple resonant scatterings with H\,{\sc i}. It serves as a critical diagnostic for the interplay between gas, dust, and radiation in star-forming galaxies and is directly linked to the ISM geometry, kinematics, and feedback processes \citep[e.g.,][]{Hayes2011, Henry2015, Runnholm2020, Melinder2023}. Spatially resolved measurements of $\rm f_{\rm esc}^{\rm Ly\alpha}$ thus provide valuable insight into the local conditions that regulate photon escape, allowing us to identify transparent channels or regions affected by outflows.

Following \cite{Melinder2023}, we computed the escape fraction on a spaxel-by-spaxel basis as
\begin{equation}
\rm f_{\mathrm{esc}}^{\mathrm{Ly}\alpha} = \frac{F_{\mathrm{obs}}^{\mathrm{Ly}\alpha}}{8.7 \times F_{\mathrm{corr}}^{\mathrm{H}\alpha}},
\end{equation}

\noindent where $\rm F_{\mathrm{obs}}^{\mathrm{Ly}\alpha}$ is the observed Ly$\rm \alpha$ flux and $\rm F_{\mathrm{corr}}^{\mathrm{H}\alpha}$ is the dust-corrected H$\rm \alpha$ flux derived from the Balmer decrement.

The resulting spatially resolved $\rm f_{\rm esc}^{\rm Ly\alpha}$ map is shown in Figure~\ref{f_esc} (left). We note that the escape fraction derived here represents the observed Ly$\alpha$ escape fraction, which reflects the combined effects of Ly$\alpha$ escape through the ISM and subsequent transmission through the inter-galactic medium (IGM). Consequently, the measured $\rm f_{\rm esc}^{\rm Ly\alpha}$ should be interpreted as an effective escape fraction rather than a purely ISM escape fraction. Our zELDA II modeling (see Appendix~\ref{app:zelda}), which explicitly accounts for IGM attenuation, yields a global Ly$\alpha$ escape fraction broadly consistent with the measured $\rm f_{\rm esc}^{\rm Ly\alpha}$.

To show the distribution of the escape fraction, in Figure~\ref{f_esc} (right) we also plot the radial profiles along the slit that is shown in the left image. The plot shows the variation of the observed ${\rm Ly\alpha}$/H${\rm \alpha}$ ratio (blue curve) and the derived ${\rm Ly\alpha}$ escape fraction (red curve) as a function of distance along the PA= 127$^\circ$ strip, oriented as the disk. The shaded regions represent the associated $\rm 1\sigma$ uncertainty.

Across the central regions, the 
observed ${\rm Ly\alpha}$/H${\rm \alpha}$ ratio remains lower, implying substantial ${\rm Ly\alpha}$ attenuation, likely due to resonant scattering and dust absorption. The escape fraction also follows a similar trend, showing a minimum near central spaxel where ${\rm Ly\alpha}$ suppression is the strongest that suggests enhanced scattering or higher dust column density. A sharp increase in both Ly$\alpha$/H$\alpha$ and $\rm f_{\rm esc}^{\rm Ly\alpha}$ is observed near Clump~A, suggesting more efficient escape of Ly$\alpha$ photons. This interpretation is consistent with the zELDA fit to the Ly$\alpha$ profile of Clump~A (see Appendix~\ref{app:zelda}). Although the uncertainties remain large, the inferred $\rm f_{\rm esc}^{\rm Ly\alpha}$ ($\sim27{-}87\%$) is consistent with enhanced Ly$\alpha$ escape from this region. 

While changes in the $\rm f_{\rm esc}^{\rm Ly\alpha}$ could be facilitated by several processes, for instance outflows driven by AGNs or star formation, reduced dust attenuation, or anisotropic \ion{H}{I} geometry, where Ly$\alpha$ photons escape preferentially along lower-opacity channels. Even though the global zELDA fit indicates a relatively high characteristic \ion{H}{I} column density ($\rm log~ N_{HI}\approx20.5~cm^{-2}$; see Appendix~\ref{app:zelda}), the increased extinction in Clump B provides a straightforward, plausible explanation for the change in Ly$\alpha$ escape. This is further supported by evidence of increased [C\,\textsc{ii}] brightness, which likely indicates a higher  H\,\textsc{i} gas column density \citep{HerreraCamus2025}.

\subsection{Ionization efficiency}

\begin{figure}
\
\includegraphics[width=\columnwidth, trim= 0 0cm 0 0]{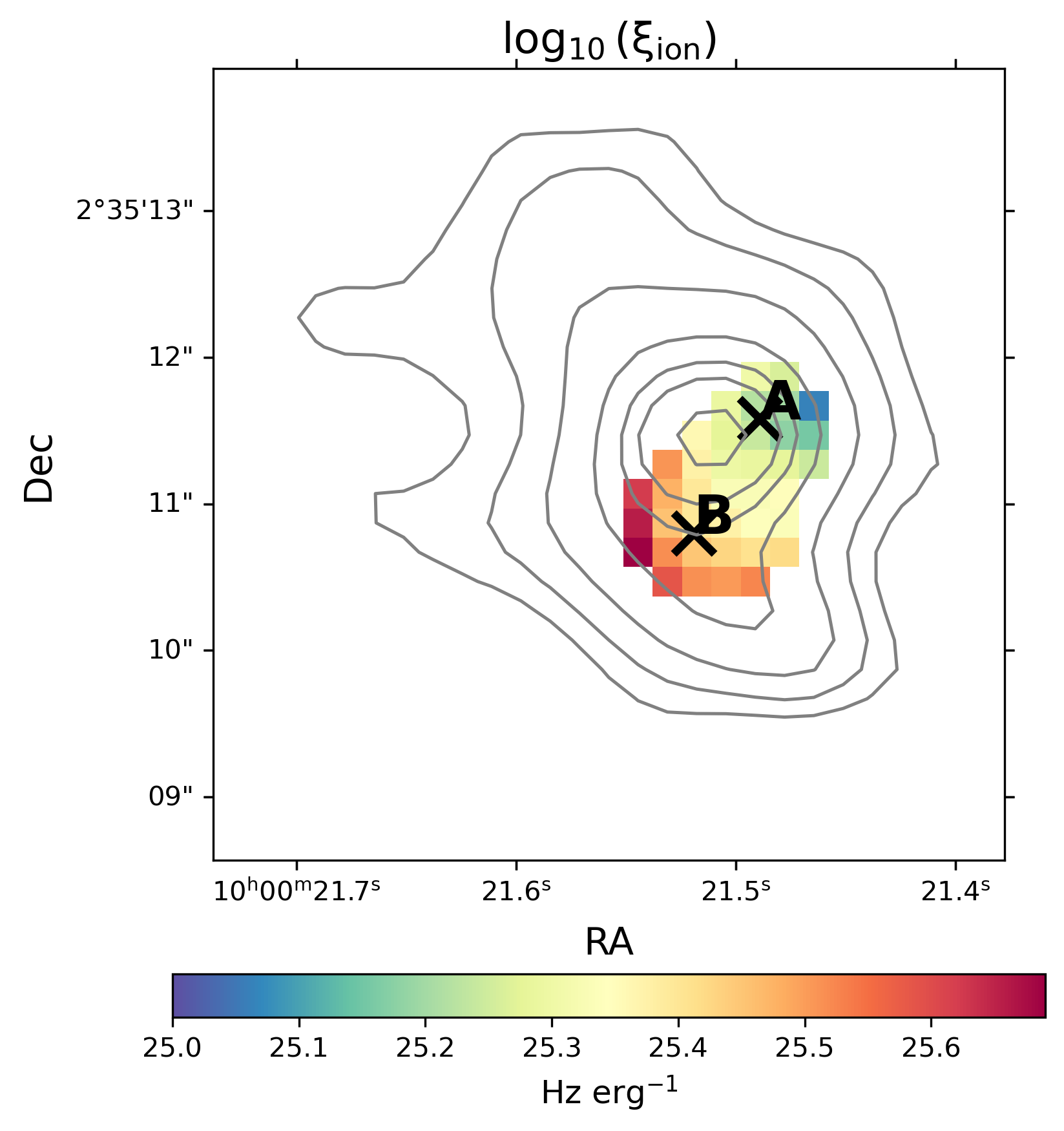}
\caption{\small Spatial map of ionization efficiency ($\rm \xi_{ion}$) with positions of two AGN candidates (A and B) marked with black crosses. Ly$\rm \alpha$ contours are overplotted in black.
}
\label{ion_eff}
\end{figure}

The ionizing efficiency, often denoted as $\xi_{\rm ion}$, quantifies the production rate of hydrogen-ionizing photons relative to the intrinsic UV luminosity of a galaxy. It is a crucial parameter for understanding the role of galaxies in cosmic reionization, as it links the observed UV continuum to the number of photons capable of ionizing H\,\textsc{i} ($\lambda < 912$~\AA) in the IGM \citep[e.g.,][]{Bouwens2016, Dayal2018, Robertson2013}. The ionizing efficiency can be expressed as

\begin{equation}
\rm \xi_{\rm ion} = \frac{Q_{\rm ion}}{L_{\nu, \rm UV}} \quad [\mathrm{photons~erg^{-1}~Hz}],
\end{equation}

\noindent where $\rm L_{\nu, UV}$ is the dust-corrected UV luminosity per unit frequency, obtained by converting the H$\alpha$ attenuation to UV attenuation assuming the \cite{Calzetti2000} law ($\rm A_{UV} = 3.1 \times A_{H\alpha}$) and applying this correction to the observed UV flux. The term $\rm Q_{\rm ion}$ is the production rate of Lyman-continuum photons, which is calculated from the intrinsic H$\rm \alpha$ luminosity \citep{Osterbrock1989} as

\begin{equation}
\rm Q_{\rm ion} = c_{\rm H\alpha} \, \frac{L_{\rm H\alpha, intr}}{(1 - f_{\rm esc}^{LyC})}, \quad 
c_{\rm H\alpha} = 7.35 \times 10^{11}~\mathrm{erg^{-1}} \ ,
\end{equation}

\noindent where $\rm f_{esc}^{LyC}$ is the Lyman-continuum escape fraction. As the $\rm f_{esc}^{LyC}$ cannot be measured directly at $z \sim 5$, we assumed $f_{\rm esc}^{\rm LyC}=0$, implying that all ionizing photons are absorbed within the galaxy and converted into nebular recombination emission under Case B recombination. Consequently, the derived $\rm Q_{\rm ion}$ represents the intrinsic ionizing photon production rate under this assumption. If LyC photons do escape, the estimated $\rm Q_{\rm ion}$ would be a lower limit. The intrinsic H$\rm \alpha$ luminosity was derived from single-Gaussian fits to the emission line and corrected for dust attenuation following \citet{Calzetti2000}. Although this approach does not explicitly separate a potential broad component from the Type-1 AGN, the broad-line region is spatially unresolved at the MUSE resolution and is expected to be confined to the nucleus. As our analysis focuses on spatially extended clumps, the measured H$\rm \alpha$ luminosities are therefore dominated by narrow emission such that the use of a single-Gaussian fit does not significantly affect our results. So we calculated the intrinsic H$\alpha$ luminosity as $\rm L_{\rm H\alpha, intr} = L_{\rm H\alpha, obs} \times 10^{\rm 0.4E(B-V)_{\rm gas}  k_{\rm H\alpha}}$, where $\rm E(B-V)_{\rm gas}$ is the color excess for the gas and $\rm k_{\rm H\alpha}$ is the extinction coefficient at H$\rm \alpha$.

For CRISTAL-02, we obtained the average value as $\log_{10}(\xi_{\rm ion}/\rm Hz~erg^{-1}) \approx 25.20^{+0.18}_{-0.03}$. In Figure~\ref{ion_eff}, we present the spatial map of the ionizing photon production efficiency. Clump~B shows a $\sim0.2$ dex higher value ($\approx 25.29^{+0.06}_{-0.09}$) compared to Clump~A ($\approx 25.07^{+0.13}_{-0.08}$), although the measurements remain consistent within uncertainties. The difference may indicate a modestly higher ionization efficiency in Clump~B, potentially linked to young stellar populations or contributions from AGN-related photoionization. Both clumps lie within the typical range reported for high-redshift ($z \sim 4$--5) star-forming galaxies \citep[e.g.,][]{Bouwens2016,Faisst2019,Robertson2022}, although more extreme systems can reach higher efficiencies.


\subsection{Nature of Ly$\rm \alpha$ emission}
\label{lya_nature}

The Ly$\alpha$ emission in CRISTAL-02 exhibits interesting spatial and kinematic behavior compared to the non-resonant tracers H$\alpha$ and the UV continuum, highlighting the dominant role of radiative transfer effects in shaping its observed properties. While the H$\alpha$ and UV emission primarily trace the star-forming disk, the Ly$\alpha$ emission is significantly more extended and shows a projected spatial alignment with the [C\,\textsc{ii}] outflow (see Figure~\ref{lyaHa_fluxmap}, first panel). Regarding the geometry of the Ly$\alpha$ emission with respect to [C\,\textsc{ii}] emission, we suspect that the [C\,\textsc{ii}] outflow may be oriented such that it propagates through the northern region while originating from the southern side of the system. Supporting this interpretation, we detected a faint southern extension of Ly$\alpha$ emission at about $\sim2''$ from the center, which may trace a weaker component of the outflowing gas. This interpretation is consistent with a scenario in which a substantial fraction of Ly$\alpha$ photons do not escape directly from their production sites but are instead scattered or reprocessed by outflowing H\,\textsc{i} gas, thereby tracing the distribution and kinematics of the surrounding gaseous environment.
However, we note that the interpretation of Ly$\alpha$ kinematics is not straightforward because the observed Ly$\alpha$ velocity field is strongly influenced by radiative transfer effects, including scattering geometry and H\,\textsc{i} column density (see Appendix~\ref{app:zelda}). The projected spatial alignment between the Ly$\alpha$ emission and the [C\,\textsc{ii}] outflow is suggestive of a possible connection, consistent with scenarios in which feedback-driven outflows may create low-opacity channels that facilitate the escape of Ly$\alpha$ photons \citep[e.g.,][]{Verhamme2006, Chonis2013, Yang2016}. A spatially resolved comparison of Ly$\alpha$ and [C\,\textsc{ii}] velocity offsets was presented in \cite{Guaita2022}, although the available S/N did not permit any robust conclusions regarding spatial variations in the Ly$\alpha$ kinematics.

The resolved Ly$\alpha$/H$\alpha$ and Ly$\alpha$/UV ratio maps further support this picture. Regions associated with Clump A exhibit enhanced Ly$\alpha$/H$\alpha$ ratios and therefore higher Ly$\alpha$ escape fractions that are coincident with lower dust attenuation inferred from the Balmer decrement. In contrast, Clump B shows strong H$\alpha$ and UV emission but suppressed Ly$\alpha$, despite a higher ionizing efficiency. This combination indicates that local ionizing photon production alone is insufficient to guarantee Ly$\alpha$ visibility and that the column density, geometry, and kinematics of H\,\textsc{i} gas play a crucial role in regulating Ly$\alpha$ escape.

The observed properties of CRISTAL-02 are broadly consistent with those of extended Ly$\alpha$-emitting systems. For instance, a $\sim40$ kpc Ly$\alpha$ blob at $z \sim 4.54$ shows spatially extended emission with a morphology that is not fully aligned with the underlying stellar and nebular components \citep{Jimnez2023}. Similar spatial offsets are also reported in proto-cluster environments at $z \sim 4.3$, where the extended Ly$\alpha$ emission is likely influenced by scattering or redistribution of ionizing photons \citep{Apostolovski2024}. These results suggest that extended Ly$\alpha$ emission can trace gas affected by radiative transfer and feedback processes, rather than arising solely from in-situ recombination.

Recent JWST/NIRSpec observations provide further constraints on the connection between Ly$\alpha$ emission and ionizing photon escape. Studies of Ly$\alpha$ emitters such as CR7 at $z = 6.6$ have used Ly$\alpha$/H$\alpha$ measurements to estimate Lyman-continuum escape fractions \citep{Marconcini2025}, while stacking analysis has revealed spatial variations in the escape fraction on subgalactic scales \citep{Tripodi2026}. Together, such analyses highlight the utility of Ly$\alpha$-based diagnostics in probing both global and spatially resolved escape processes. In this context, our analysis of CRISTAL-02 extends these approaches by examining the spatially resolved Ly$\alpha$–H$\alpha$ connection.

\subsection{Caveats on AGN contribution and interpretation of derived quantities}

The presence of AGN activity in CRISTAL-02 remains uncertain. Evidence of a possible AGN primarily comes from the detection of broad H$\alpha$ components reported in the recent JWST/NIRSpec study of the CRISTAL sample by \citet{Ren2025}. Clump A exhibits a broad H$\alpha$ component with ${\rm FWHM}=1618~{\rm km~s^{-1}}$, while Clump B shows a broad H$\alpha$ component with ${\rm FWHM}=955~{\rm km~s^{-1}}$. Furthermore, the authors performed a detailed emission-line diagnostic analysis (see Section~5.4 of \citealt{Ren2025}), finding that the two AGN candidates lie close to the boundary between the star-forming and AGN populations and occupy the composite region in most diagnostic diagrams. They concluded that none of the available emission-line diagnostics provides definitive evidence of AGN activity and interpreted the observed ambiguity as consistent with a weak AGN embedded within a vigorously star-forming system.

Recent work by \citet{Davies2026} provides additional context for the nature of the outflow in CRISTAL-02. Combining JWST/NIRSpec and ALMA observations, they identified a multiphase outflow associated with the region they refer to as Clump B, which corresponds to Clump D in the nomenclature of \citet{Ren2025}. This region is not separately classified as a distinct clump in our analysis. 
As discussed in Section~\ref{previous_study}, the limited spatial resolution of our data may blend emission from the potential AGN candidate and the outflow in the southern region. \citet{Davies2026} measured outflow velocities of $v_{\rm out}({\rm H}\alpha)=550^{+80}_{-70}\ {\rm km~s^{-1}}$, $v_{\rm out}([{\rm O\,\textsc{iii}}])=740^{+150}_{-120}\ {\rm km~s^{-1}}$, and $v_{\rm out}([{\rm C\,\textsc{ii}}])=640^{+70}_{-60}\ {\rm km~s^{-1}}$. They derived a mass-loading factor of $\eta=2.0\pm0.4$, indicating that the outflow rate is approximately twice the current star formation rate. Importantly, they showed that the inferred outflow kinetic power ($\rm \dot{E}_{\rm out}=6.6\pm1.4\times10^{43}\ {\rm erg~s^{-1}}$) is comparable to the energy injection rate from supernovae ($\rm \dot{E}_{\rm SN}=6.2\pm1.1\times10^{43}\ {\rm erg~s^{-1}}$). The measured outflow velocity is consistent with expectations from scaling relations based on the star formation rate surface density, while the comparable energy injection rates indicate that stellar feedback is energetically sufficient to power the observed outflow. Combined with the lack of X-ray and radio detections, a multiwavelength spectral energy distribution consistent with star formation, and emission-line ratios consistent with stellar photoionization, these findings led \citet{Davies2026} to conclude that stellar feedback is the most likely primary driver of the outflow, although a contribution from AGN activity cannot be completely ruled out.

Alternative AGN diagnostics remain similarly inconclusive. For example, additional constraints on AGN activity can be obtained from the [O\,\textsc{iii}]$\lambda5007$/[O\,\textsc{iii}]$\lambda4363$ and [O\,\textsc{iii}]$\lambda4363$/H$\gamma$ line ratios \citep[e.g.,][]{Mazzolari2024, Torralba2024}. To investigate this possibility, \citet{Ren2025} searched for auroral [O\,\textsc{iii}]$\lambda4363$ emission in both CRISTAL-02 AGN candidates. However, only $3\sigma$ upper limits were obtained for [O\,\textsc{iii}]$\lambda4363$ ($<0.36\times10^{-18}$ and $<0.15\times10^{-18}~{\rm erg~s^{-1}~cm^{-2}}$ for Clumps A and B, respectively), preventing robust constraints from either the [O\,\textsc{iii}]$\lambda5007$/[O\,\textsc{iii}]$\lambda4363$ or [O\,\textsc{iii}]$\lambda4363$/H$\gamma$ diagnostics. Other observational properties also do not strongly support a dominant AGN contribution. For instance, the modest dust attenuation that we observed in Fig.~\ref{ebv_Av_maps} ($\rm E(B-V)\sim$0 - 0.25) suggests that a luminous, heavily obscured AGN is unlikely to be energetically dominating the ionized gas emission.

We note that the present work is not intended to provide an independent AGN classification. Our analysis uses the same JWST/NIRSpec data analyzed by \citet{Ren2025} and \citet{Davies2026}; however, we convolved the emission-line maps to the MUSE spatial resolution to enable a direct spatial comparison between the Ly$\alpha$ and H$\alpha$ emission. This PSF matching smooths compact structures and blends emission from neighboring regions, preventing a robust decomposition of broad and narrow emission-line components and limiting the applicability of standard AGN diagnostics. Thus, our analysis does not provide independent evidence of AGN activity in CRISTAL-02. We therefore remain cautious about attributing the observed ionization and outflow properties to AGN activity. Deeper spatially resolved multiwavelength observations are required to determine whether AGN feedback plays any significant role in this system.

While our analysis assumes that the observed H$\alpha$, H$\beta$, UV, and Ly$\alpha$ emission primarily trace star formation, a possible AGN contribution introduces important uncertainties in the physical interpretation of the derived quantities. In particular, $f_{\rm esc}^{\rm Ly\alpha}$ and $\xi_{\rm ion}$ rely on the assumption of stellar photoionization and Case B recombination. If AGNs contribute significantly to the nebular or UV emission, the intrinsic Ly$\alpha$/H$\alpha$ ratio may deviate from the canonical value, and the dust-corrected H$\alpha$ luminosity may no longer trace purely stellar recombination. In this scenario, $f_{\rm esc}^{\rm Ly\alpha}$ should be regarded as an effective quantity rather than a true escape fraction. Similarly, $\xi_{\rm ion}$ may be biased because both the ionizing photon production rate and UV luminosity could include contributions from AGN emission.

Despite these uncertainties, the spatially resolved line ratios analysed in this work (e.g., Ly$\alpha$/H$\alpha$ and Ly$\alpha$/UV) remain robust observational diagnostics, as they directly trace the relative spatial distribution of the emission and are largely insensitive to the nature of the underlying ionizing source. Therefore, the observed contrasts between Clumps A and B can still be interpreted in terms of differences in gas geometry, dust attenuation, and Ly$\alpha$ radiative transfer conditions. Nevertheless, throughout this paper any discussion of AGN-regulated Ly$\alpha$ escape, ionization structure, or outflow activity should be interpreted as a plausible scenario rather than definitive evidence of AGN feedback in CRISTAL-02.


\section{Conclusions}
\label{conclusion}

We have presented a spatially resolved analysis of Ly$\alpha$, H$\alpha$, and H$\beta$ emission in CRISTAL-02 galaxy at $z \sim 5.3$, combining MUSE and JWST/NIRSpec IFU data at a matched spatial resolution. The physical conditions and morphology of CRISTAL-02 make it an ideal laboratory for spatially resolved Ly$\alpha$–H$\alpha$ comparisons. The system resides in an overdense protocluster environment and exhibits a clumpy irregular structure that is consistent with ongoing interaction or merger-driven activity. Its extended Ly$\alpha$ emission, combined with the H$\alpha$ morphology at matched spatial resolution, enables a direct comparison between resonant and non-resonant gas tracers. In addition, the presence of cold gas and dust, along with the evidence of outflows, provides a natural framework to test whether Ly$\alpha$ escape is governed by feedback-driven processes or the underlying gas distribution.

We summarize the main findings of the paper below:

\begin{enumerate}
    \item We find that Ly$\alpha$ emission is significantly more extended than H$\alpha$ and UV emission and exhibits a projected spatial alignment with the [C\,\textsc{ii}] outflow.

    \item The observed geometry suggests a complex outflow structure, with the faint southern Ly$\alpha$ extension potentially tracing a weaker outflow component. The alignment between Ly$\alpha$ emission and the [C\,\textsc{ii}] outflow further suggests that feedback-driven processes create low-opacity channels that facilitate anisotropic Ly$\alpha$ escape.

    \item The galaxy has been identified with multiple clumpy structures, but two compact regions (Clump~A and ~B) are suggested to be potential AGN candidates based on H$\rm \alpha$ broad-line analysis \citep{Ren2025}. Clump A shows lower extinction, a higher Ly$\rm \alpha$/H$\rm \alpha$, and a higher Ly$\rm \alpha$ escape fraction, whereas Clump B is dustier and H$\rm \alpha$ bright and exhibits suppressed Ly$\rm \alpha$ despite a higher ionization efficiency.

    \item While the presence of an AGN in Clumps A and B cannot be ruled out, the observed Ly$\alpha$ morphology, spatial variations in Ly$\alpha$ escape, and velocity offsets can be consistently explained within a framework where outflows and local ISM conditions regulate Ly$\alpha$ radiative transfer. This interpretation is further supported by the \texttt{zELDA II} radiative transfer modeling, which favors a low-velocity outflow together with a relatively high H\,\textsc{i} column density, in the direction of the observed extended Ly$\alpha$ emission. The recent results by \citet{Davies2026} favor a star formation-driven origin for the outflow. However, the available data do not yet uniquely distinguish between stellar and AGN-driven feedback.
\end{enumerate}

Overall, this study highlights the importance of spatially resolved analyses in understanding the interplay between star formation, feedback, and radiative transfer in high-z galaxies.
Further insight into the nature of the Ly$\alpha$ emission in CRISTAL-02 would benefit from detailed radiative transfer modeling of the Ly$\alpha$ line profiles in combination with modeling of the spatial distribution \citep[e.g.,][]{Maselli2005, Stern2021, Smith2025}. Such modeling could constrain the H\,\textsc{i} column density, velocity field, and geometric configuration of the gas and place quantitative limits on the role of outflows and clumpiness in shaping the Ly$\alpha$ escape. Future studies incorporating additional non-resonant tracers (e.g., [O\,\textsc{iii}] $\lambda$5007) would provide tighter constraints on the ionization conditions in the galaxy. Complementary observations with higher-S/N Ly$\alpha$ spectroscopy, along with spatially resolved measurements across a broader sample, will be essential to placing CRISTAL-02 in a wider evolutionary context and for testing whether similar mechanisms regulate Ly$\alpha$ escape in other high-z galaxies.

\begin{acknowledgements}
    We thank the anonymous referee for their constructive comments and suggestions, which helped improve the manuscript. AA, LG and DO thank Siddhartha for his helpful suggestions with the zELDA modeling. AA and LG gratefully acknowledges financial support from ANID - MILENIO - NCN2024\_112, ANID BASAL project FB210003, and FONDECYT regular project number 1230591. MA is supported by FONDECYT grant number 1252054 and gratefully acknowledges support from the ANID Basal Project FB210003, ANID Milenio NCN2024\_112, and ANID Vinculaci\'on Internacional FOVI250261. RLD is supported by the Australian Research Council through the Discovery Early Career Researcher Award (DECRA) Fellowship DE240100136 funded by the Australian Government. V.V. acknowledges support from the Comité ESO Mixto 2024 and from the ANID BASAL project FB210003. J.M. gratefully acknowledges support from ANID MILENIO NCN2024\_112. A.N. acknowledges support from the Narodowe Centrum Nauki (NCN), Poland, through the SONATA BIS grant UMO-2020/38/E/ST9/00077. M. S. was supported by the European Research Council (ERC) under the European Union’s Horizon 2020 research and innovation programme (DistantDust, Grant agreement No. 101117541)

\end{acknowledgements}

%


\begin{appendix}

\section{Surface brightness profiles}
\begin{figure}
\centering
\includegraphics[width=\columnwidth, trim= 0 0cm 0 0]{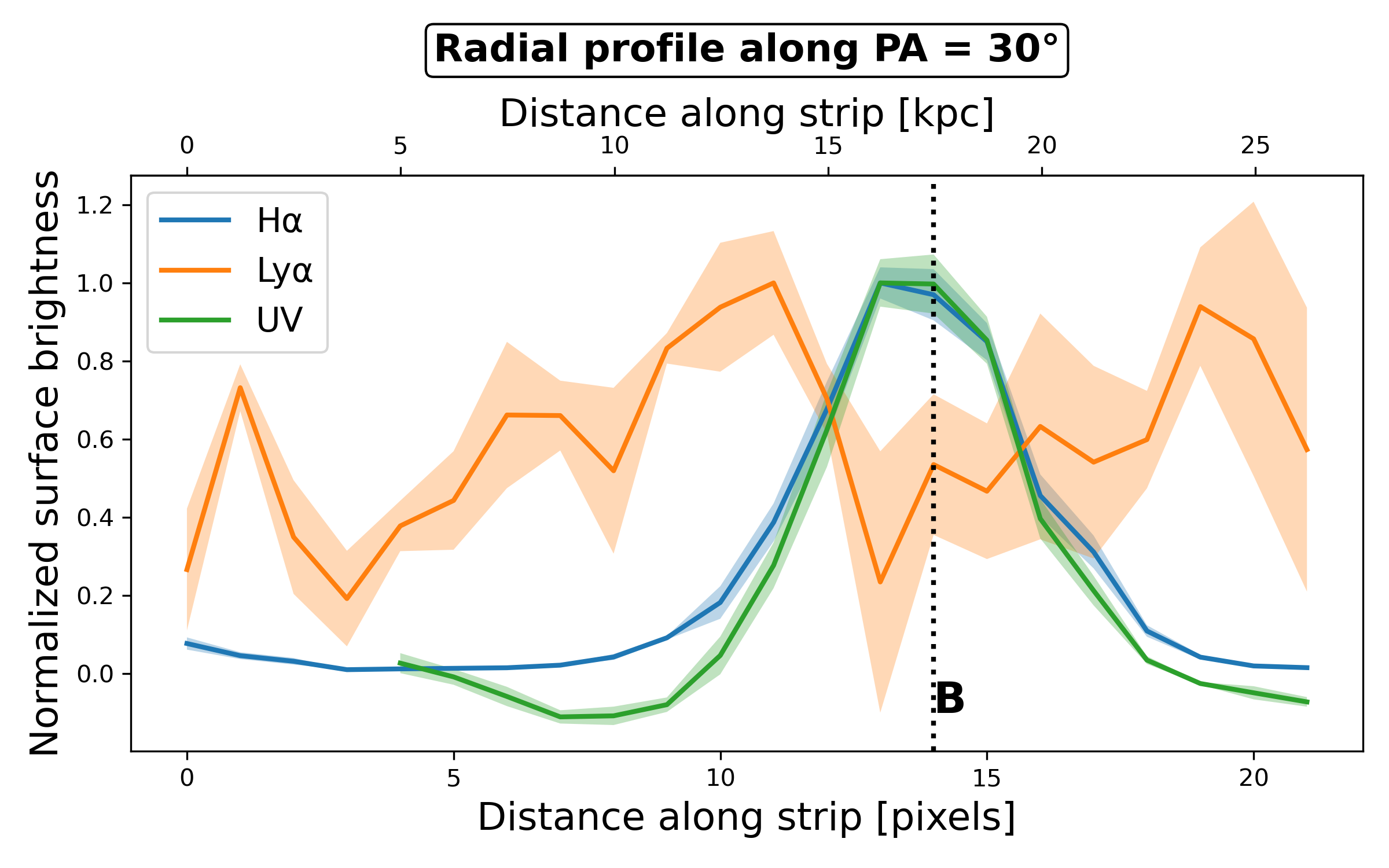}
\includegraphics[width=\columnwidth, trim= 0 0cm 0 0]{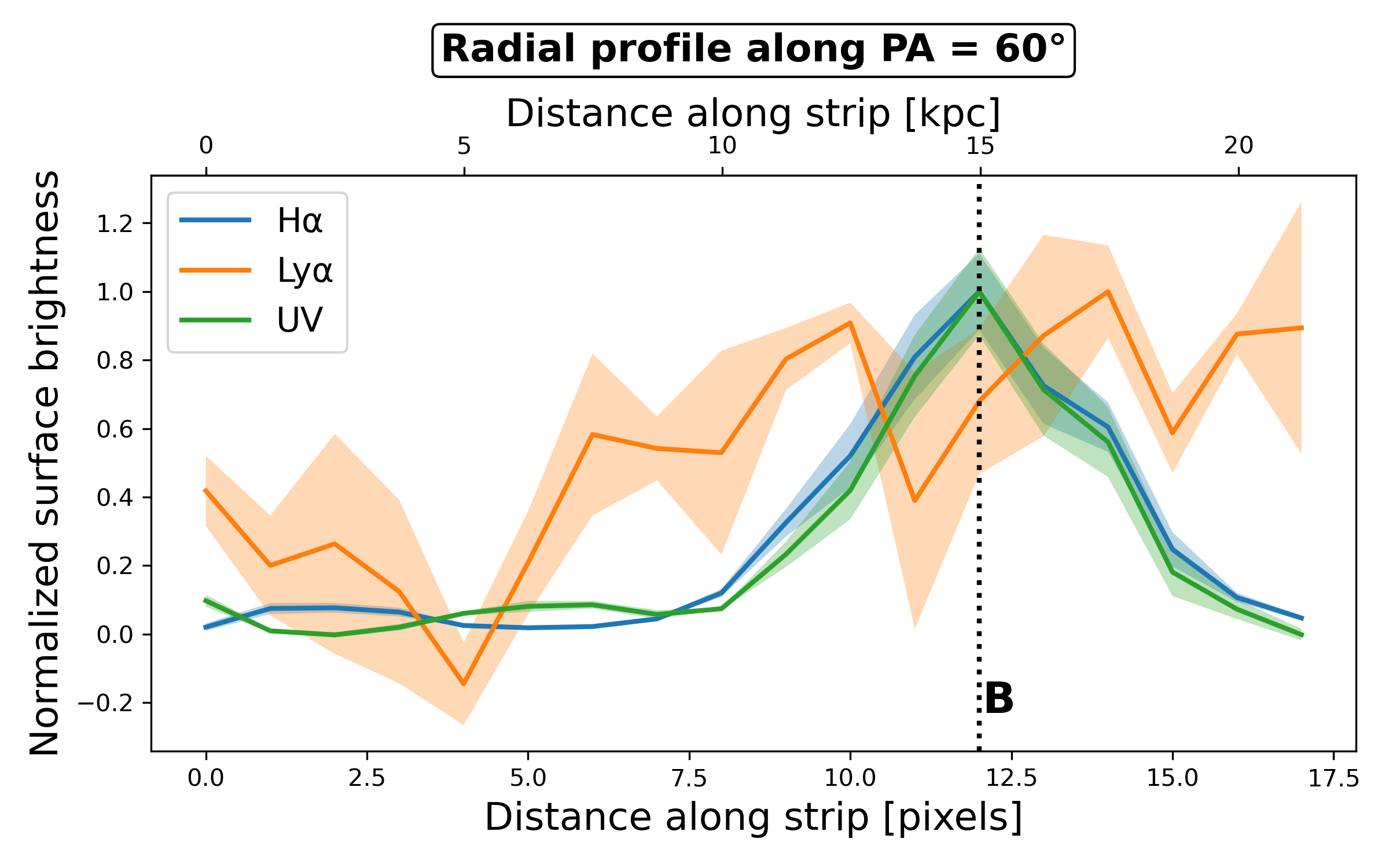}
\includegraphics[width=\columnwidth, trim= 0 0cm 0 0]{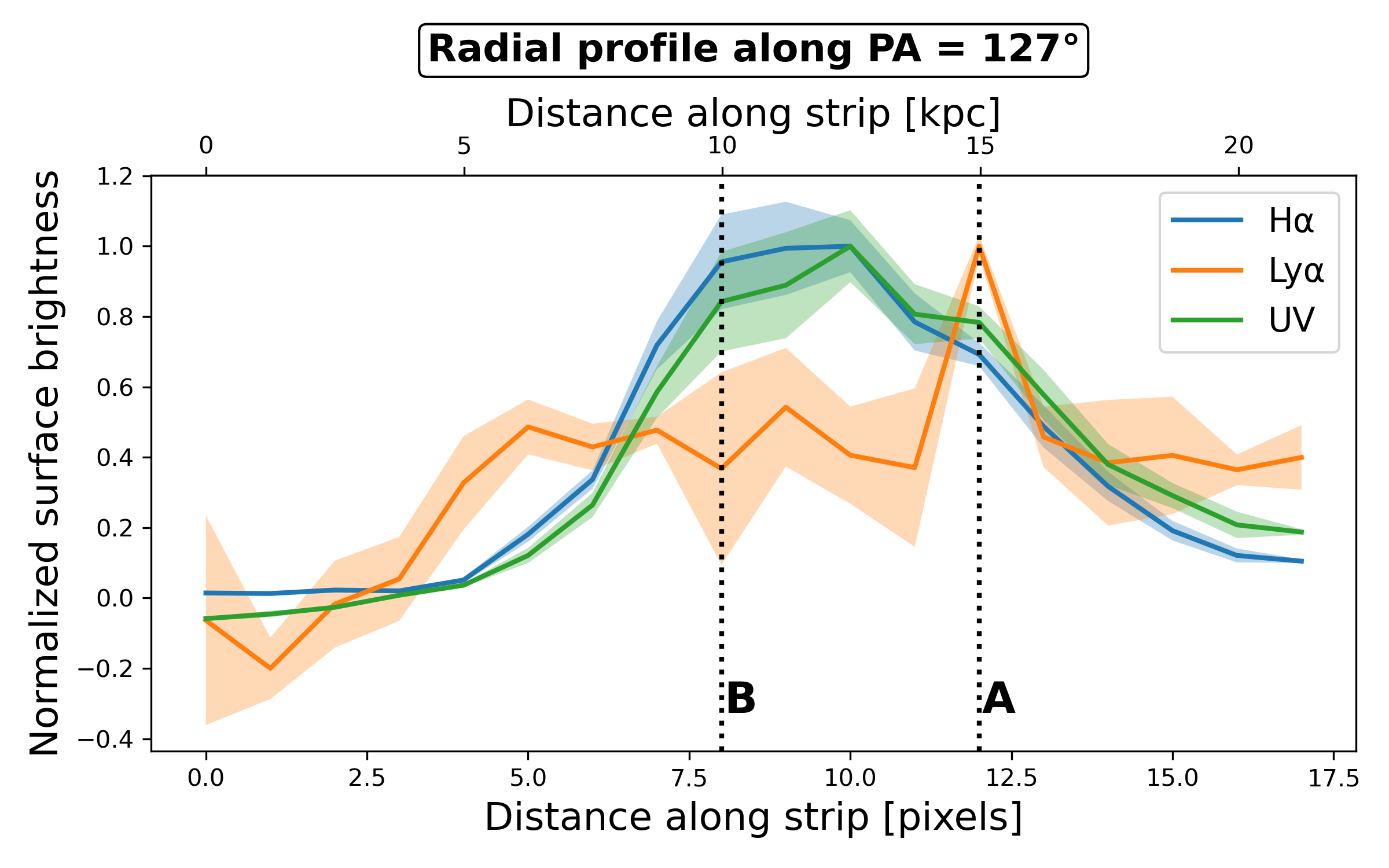}

\caption{\small 
Normalized radial surface-brightness profiles extracted along linear strips oriented at position angles PA = 30$^\circ$, 60$^\circ$, and  127$^\circ$. The profiles correspond to the H$\rm \alpha$ (blue), Ly$\rm \alpha$ (orange), and UV (green). Shaded regions indicate the 1$\sigma$ uncertainties estimated from the standard error of the mean across the strip width. Vertical dotted lines denote the projected locations of the Clump A and B.}
\label{radial_profile}
\end{figure}

We obtained the surface brightness profiles across the slits (Figure~\ref{radial_profile}). For all slits, we observe a peak in both H$\alpha$ and UV emission around the region corresponding to Clump B, indicating enhanced star-forming activity and strong ionizing photon production. Interestingly, at the same spatial position, the Ly$\alpha$ profile exhibits a dip. This suggests that, despite the presence of ionizing sources, Ly$\alpha$ photons are being efficiently absorbed or scattered out of the line of sight, likely due to a higher local H\,\textsc{i} column density and/or increased dust content within Clump B.

The surface brightness profile along PA = 127$^\circ$ shows a sharp peak in Ly$\alpha$ emission at the location of Clump A, while both H$\alpha$ and UV emission display a declining trend. This spatial offset between Ly$\alpha$ and the non-resonant tracers indicates that the Ly$\alpha$ emission at this position is not directly associated with in-situ star formation. Instead, it likely arises from resonant scattering of Ly$\alpha$ photons produced in the central star-forming regions, which subsequently escape at larger radii. The enhanced Ly$\alpha$ surface brightness at Clump A therefore suggests the presence of a preferential escape channel, potentially facilitated by a lower H\,\textsc{i} column density, anisotropic gas distribution, or outflow-driven kinematics. This behavior is consistent with scenarios in which Ly$\alpha$ photons undergo spatial diffusion prior to escape, resulting in extended emission that is decoupled from the underlying UV and H$\alpha$ morphology. As seen in Figure~\ref{radial_profile}, the relatively large uncertainties associated with the Ly$\alpha$ measurement, due to low surface-brightness sensitivity make it difficult to draw any definitive conclusions.

\section{zELDA modeling}
\label{app:zelda}

\begin{figure}
\centering
\includegraphics[width=\columnwidth, trim= 0 0cm 0 0]{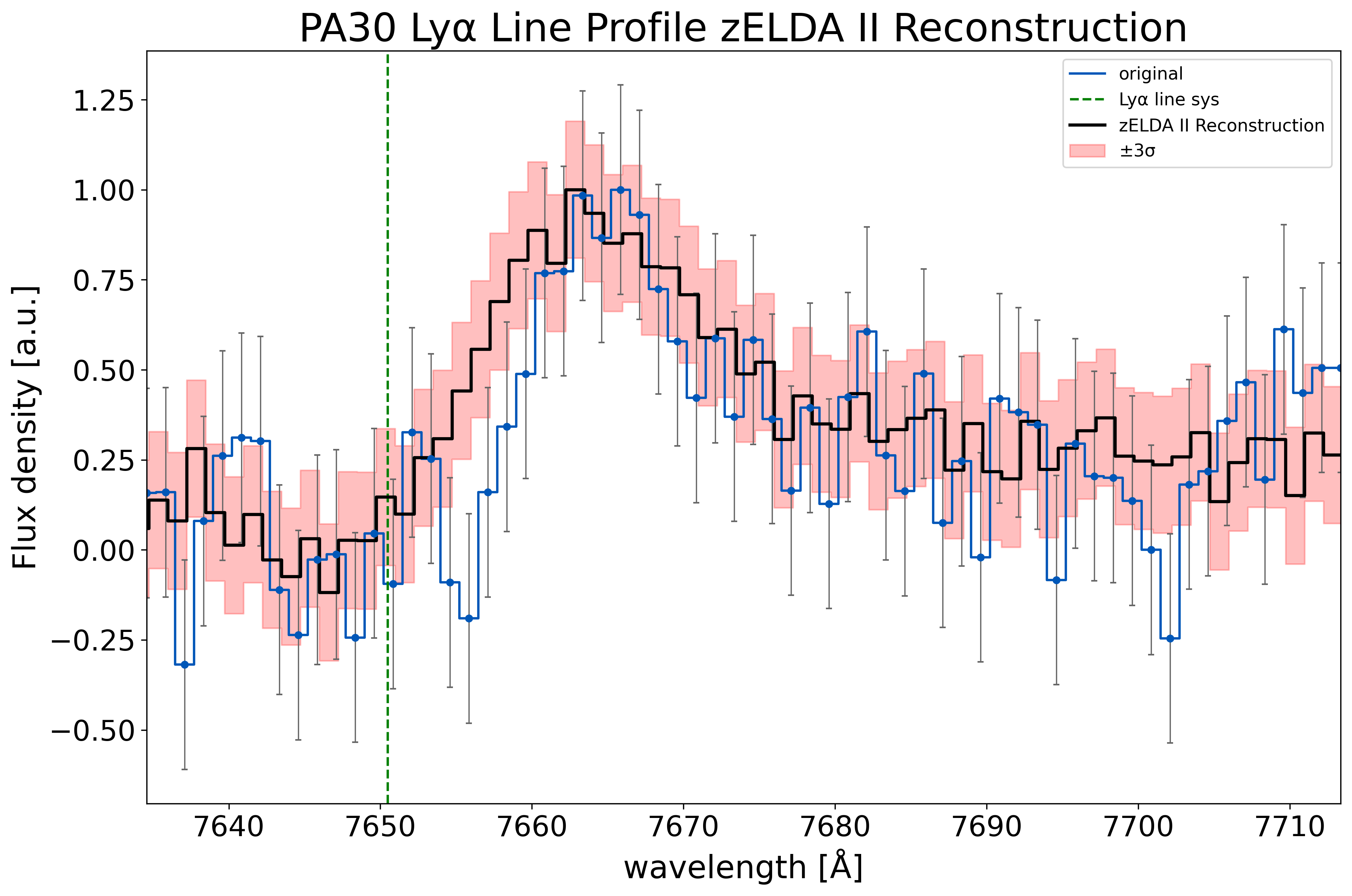}
\includegraphics[width=\columnwidth, trim= 0 0cm 0 0]{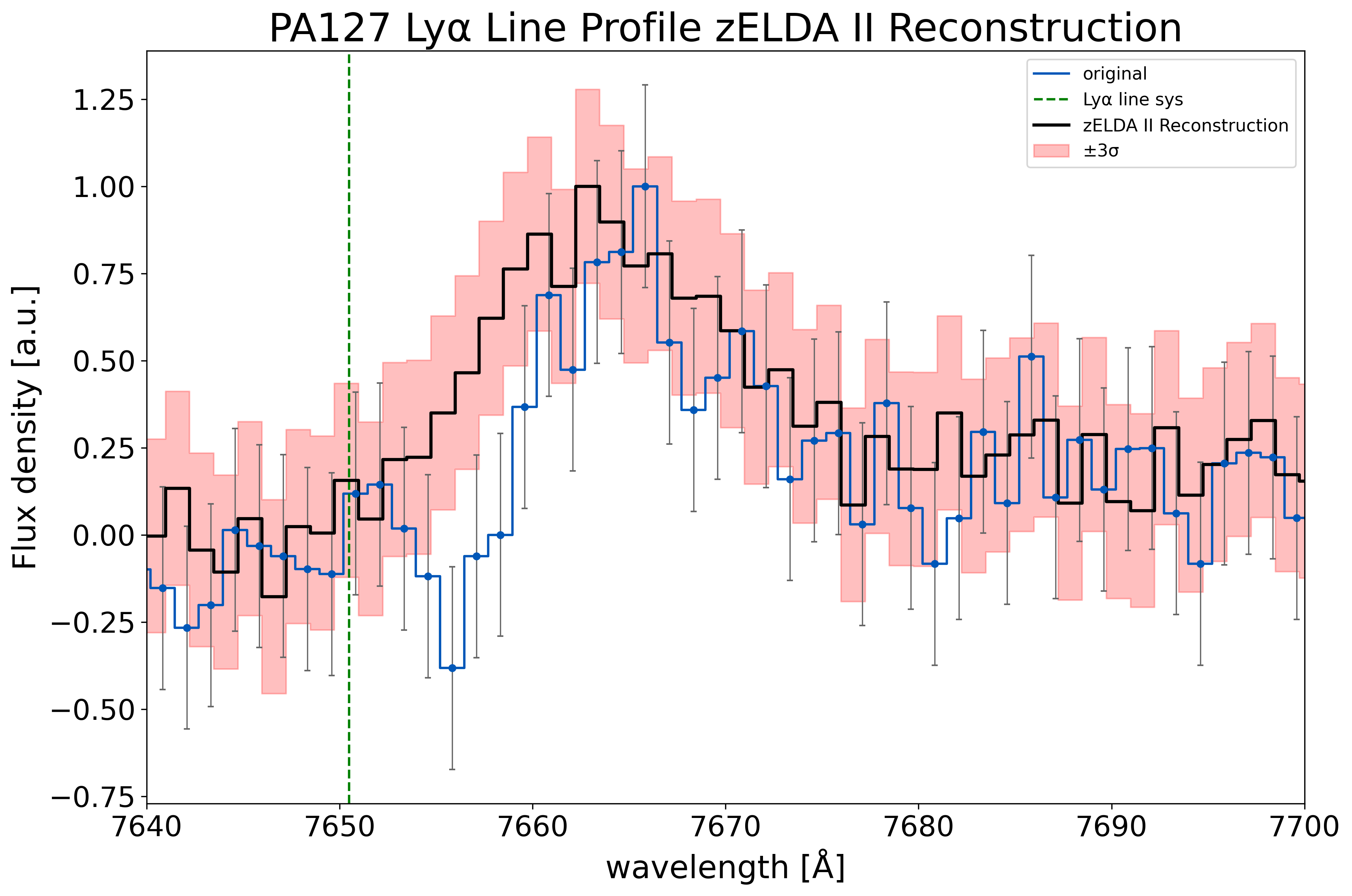}
\includegraphics[width=\columnwidth, trim= 0 0cm 0 0]{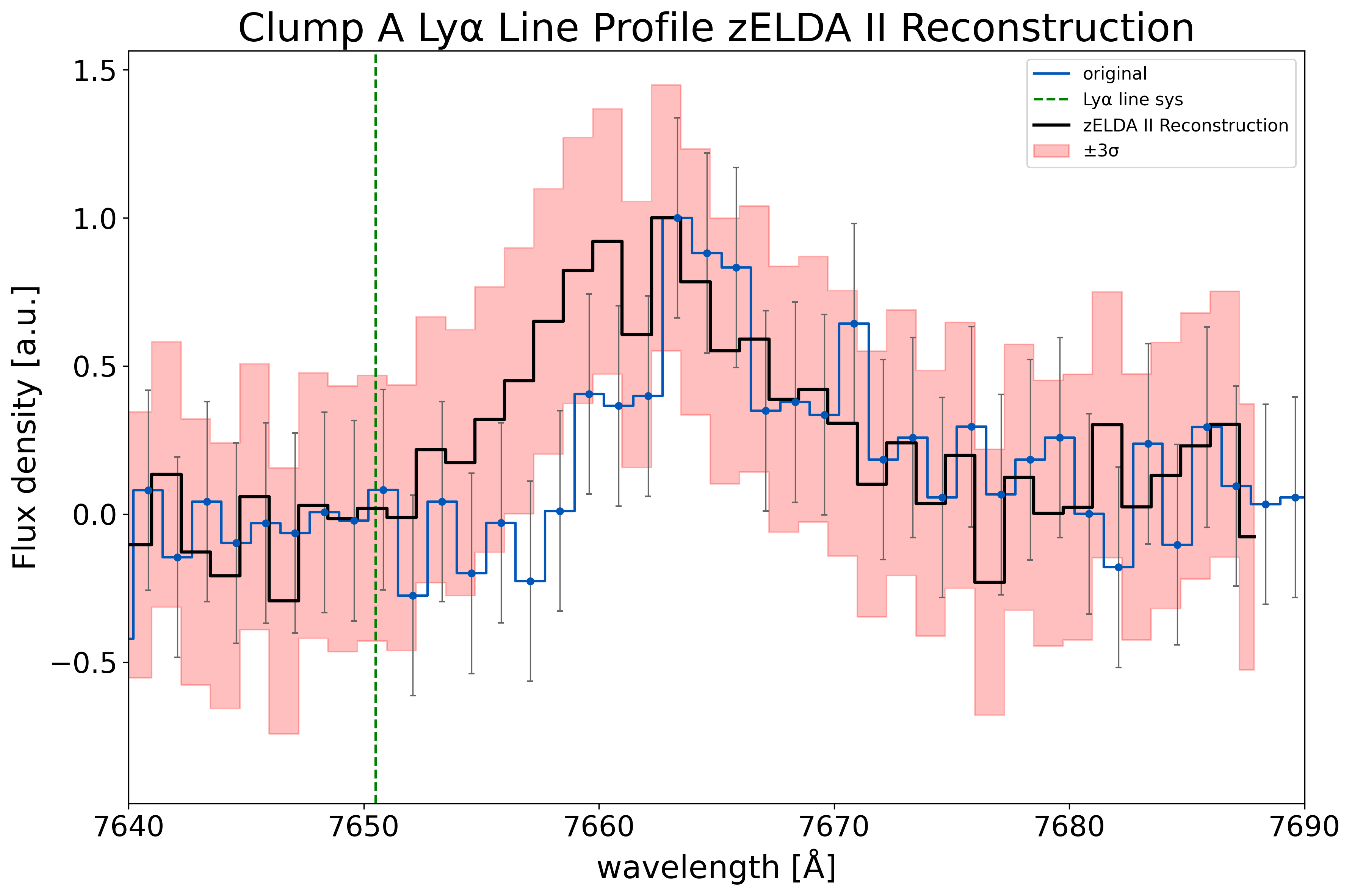}

\caption{\small Comparison between the observed Ly$\alpha$ line profile (blue) with the best-fit zELDA II reconstructed profile (black), which includes the effects of IGM transmission and is convolved to the MUSE spectral resolution. The red shaded region represents the $\rm 3\sigma$ model uncertainty, while the gray error bars indicate observational uncertainties estimated from the continuum noise. The vertical green dashed line marks the rest-frame Ly$\alpha$ wavelength at $z = 5.2931$. Top: The spectra along the slit at PA = $30^\circ$ (magenta slit in Figure~\ref{lyaHa_fluxmap}). Middle: The spectra along the slit at PA = $127^\circ$, corresponding to the spatial direction where $\rm f_{\rm esc}^{\rm Ly\alpha}$ was measured (red slit in Figure~\ref{f_esc}) Bottom: The spectra extracted from Clump A.}
\label{zelda_profile}
\end{figure}

To understand the Ly$\alpha$ kinematics along the magenta slit aligned with the [C\,\textsc{ii}] outflow opening angle (PA = 30$^\circ$, see Figure~\ref{lyaHa_fluxmap}), we modeled the extracted Ly$\alpha$ spectrum using \texttt{zELDA II} \citep{Gurung2025}, a publicly available Python package capable of fitting observed Ly$\alpha$ spectra and predicting Ly$\alpha$ line profiles through the use of neural networks (Figure~\ref{zelda_profile}). It also accounts for the impact of the IGM as a function of redshift on the resulting profile and allows for modeling at spectral resolutions comparable to that of MUSE.

We first input the spectrum corresponding to the magenta slit, which provides a set of parameters that best reproduce the observed profile. These include, among others, the outflow velocity expansion, the H\,\textsc{i} column density of the emergent profile, and the systemic redshift. Once these parameters are obtained, the Ly$\alpha$ profile can be reconstructed by generating a model using the best-fit values. In this reconstruction, \texttt{zELDA II} incorporates the effects of the IGM, enabling a direct and robust comparison between the modeled and observed profiles and allowing for a more comprehensive analysis of the Ly$\alpha$-emitters properties. The best-fit parameters derived from the modeling are summarized in Table~\ref{tab:zelda_results}. We obtained a Ly$\alpha$ line profile consistent with a low-velocity outflow velocity ($\sim 50$--$190)\ \mathrm{km\ s^{-1}}$ and an H\,\textsc{i} column density of $\sim 10^{20.5}\ \mathrm{cm^{-2}}$.

To compare the $\rm f_{\rm esc}^{\rm Ly\alpha}$ derived from the H$\alpha$-based analysis in Section~\ref{esc_fraction} with the \texttt{zELDA II} predictions, we performed an additional fitting on a Ly$\alpha$ spectrum extracted along PA = 127$^\circ$ (Figure~\ref{zelda_profile}, middle panel), corresponding to the direction where the H$\alpha$-based $\rm f_{\rm esc}^{\rm Ly\alpha}$ was measured (see Figure~\ref{f_esc}). For the best-fit solution, we obtain an ISM escape fraction of ($f_{\rm esc}^{\rm ISM}=0.73^{+0.21}_{-0.69}$) and an IGM transmission of ($f_{\rm esc}^{\rm IGM}=0.56^{+0.21}_{-0.54}$). Combining these contributions yields a global Ly$\alpha$ escape fraction of ($\sim2{-}80\%$). While the uncertainty remains large, the inferred range is broadly consistent with the H$\alpha$-based $\rm f_{\rm esc}^{\rm Ly\alpha}$ estimate ($\sim2{-}12\%$; see Section~\ref{esc_fraction}) within the uncertainties.

We also modeled the Ly$\alpha$ spectrum extracted from Clump A (Figure~\ref{zelda_profile}, bottom panel). Owing to the lower S/N($\sim6$), the constraints are weaker than for the integrated PA = 30$^\circ$ and 127$^\circ$ spectrum, while the Ly$\alpha$ profile of Clump B does not have sufficient S/N for reliable \texttt{zELDA II} modeling. For Clump A, we obtain an ISM escape fraction of ($\rm f_{\rm esc}^{\rm ISM}=0.92^{+0.07}_{-0.28}$) and an IGM transmission of ($\rm f_{\rm esc}^{\rm IGM}=0.76^{+0.18}_{-0.40}$), corresponding to a global Ly$\alpha$ escape fraction of ($\sim27{-}87\%$). Given the relatively large uncertainties, these results do not permit a robust comparison of the Ly$\alpha$ escape properties between the individual clumps.

\begin{table*}
\centering
\caption{Best-fit parameters obtained using \texttt{zELDA II} for the three spectra analyzed: the PA = $30^\circ$ slit tracing the extended Ly$\alpha$ emission, the PA = $127^\circ$ slit corresponding to the spatial direction tracing the H$\alpha$-based $\rm f_{\rm esc}^{\rm Ly\alpha}$ measurement, and Clump A.}
\label{tab:zelda_results}
\begin{tabular}{lccc}
\hline
Parameter & PA = $30^\circ$ & PA = $127^\circ$ & Clump A \\
\hline
\\
Redshift $z$
& $5.294^{+0.001}_{-0.001}$
& $5.2971^{+0.004}_{-0.004}$
& $5.2973^{+0.003}_{-0.004}$ \\

Outflow velocity $V_{\rm exp}$ (km s$^{-1}$)
& $100^{+91}_{-48}$
& $152^{+402}_{-114}$
& $84^{+84}_{-39}$ \\

$\log(N_{\rm HI}/{\rm cm^{-2}})$
& $20.55^{+0.23}_{-0.29}$
& $20.08^{+0.25}_{-0.31}$
& $19.89^{+0.17}_{-0.46}$ \\

Dust optical depth $\tau_{\rm d}$
& $0.003^{+0.004}_{-0.001}$
& $0.0052^{+0.020}_{-0.003}$
& $0.0054^{+0.013}_{-0.004}$ \\


Intrinsic line width (\AA)
& $0.85^{+0.28}_{-0.12}$
& $1.00^{+1.08}_{-0.38}$
& $0.79^{+0.87}_{-0.34}$ \\


\hline
\end{tabular}
\end{table*}

\section{Slit-extracted velocity profiles of Ly$\alpha$ and H$\alpha$ }

\begin{figure}
\centering
\includegraphics[width=\columnwidth, trim= 0 0cm 0 0]{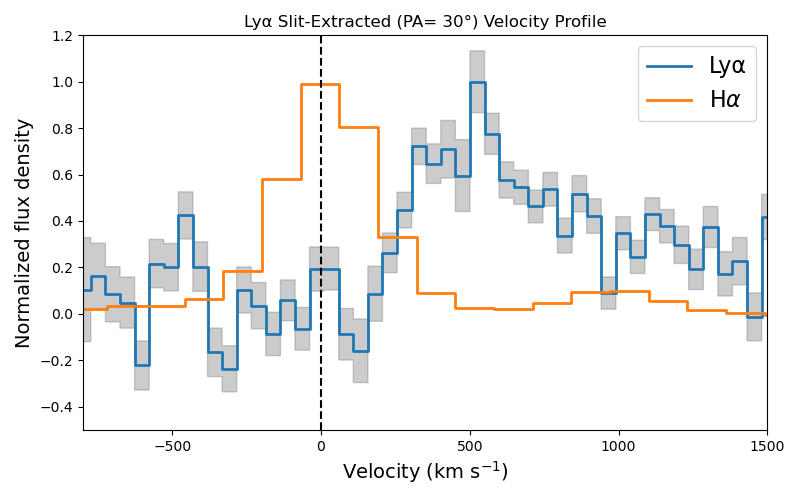}
\includegraphics[width=\columnwidth, trim= 0 0cm 0 0]{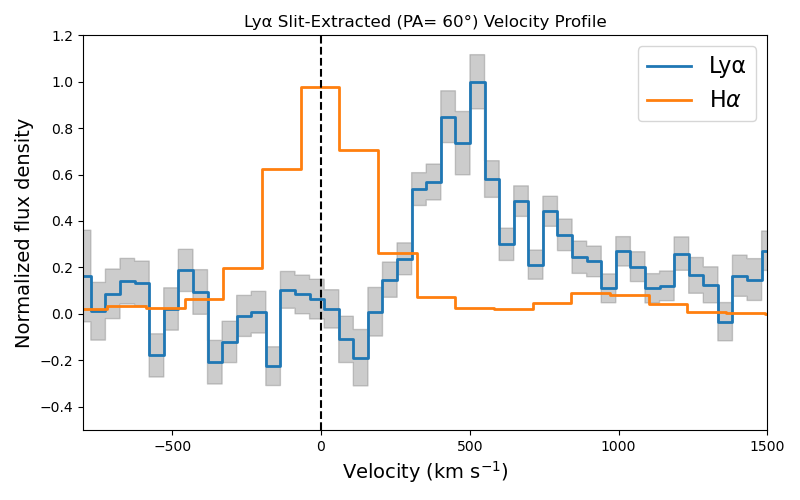}
\includegraphics[width=\columnwidth, trim= 0 0cm 0 0]{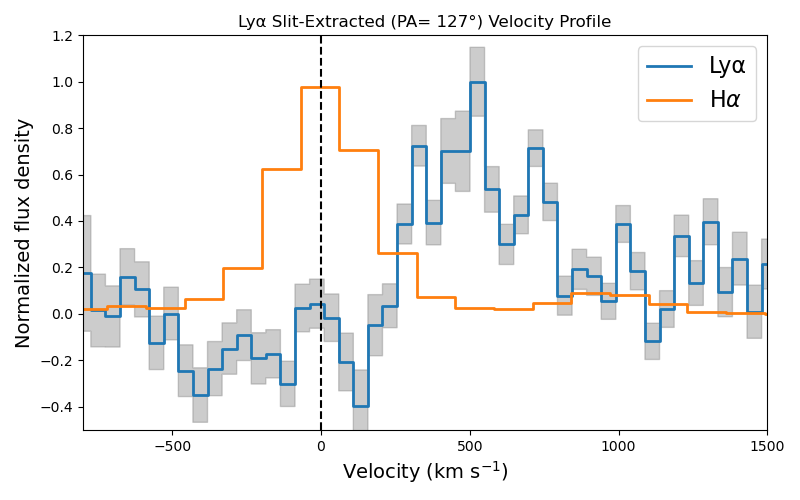}
\caption{\small 
Slit-extracted velocity profiles of Ly$\alpha$ (blue) and H$\alpha$ (orange) for a slit PA = 30$^\circ$, 60$^\circ$, and 127° respectively. Both profiles are normalized to their respective peak fluxes within the line region. The Ly$\alpha$  profile is shown with the estimated 1$\sigma$  uncertainty derived from line-free spectral channels (gray shaded region). The vertical dashed line marks the systemic velocity. }
\label{velocity_profile}
\end{figure}

In Fig.~\ref{velocity_profile}, we present the slit-extracted H$\alpha$ and Ly$\alpha$ velocity profiles relative to the systemic redshift defined by the [C\,\textsc{ii}] emission. The profiles are obtained from three independent slits oriented at position angles of $30^\circ$, $60^\circ$, and $127^\circ$. In all the three slit configurations, the Ly$\alpha$ emission exhibits broader velocity distributions than H$\alpha$. The systematically larger Ly$\alpha$ linewidth indicate that Ly$\alpha$ photons probe a more complex kinematic and radiative environment than H$\alpha$. While H$\alpha$ traces the intrinsic kinematics of ionized gas within star-forming regions, Ly$\alpha$ is a resonant transition and is therefore strongly affected by multiple scattering in H\,\textsc{i}. As a result, the Ly$\alpha$ emission is broadened and redistributed in velocity space, indicating the combined effects of gas motions, optical depth, and radiative transfer.

In addition, the Ly$\alpha$ profiles are offset from the systemic velocity, with their peaks shifted toward positive velocities relative to H$\alpha$. This velocity offset is commonly interpreted as a signature of outflowing H\,\textsc{i} gas, where Ly$\alpha$ photons preferentially escape after being back-scattered from receding gas on the far side of the system. The magnitude of this offset provides insight into the H\,\textsc{i} properties governing Ly$\alpha$ escape. In addition to outflow kinematics, a large velocity offset can arise from a high H\,\textsc{i} column density, which increases the number of resonant scatterings required before escape. This scenario is supported by our zELDA II radiative transfer modeling ($\rm \log N_{\rm HI}\approx20.5$; see Appendix~\ref{app:zelda}).





\section{Supplementary figure}

\begin{figure}
\centering
\includegraphics[width=\columnwidth, trim= 0 0cm 0 0]{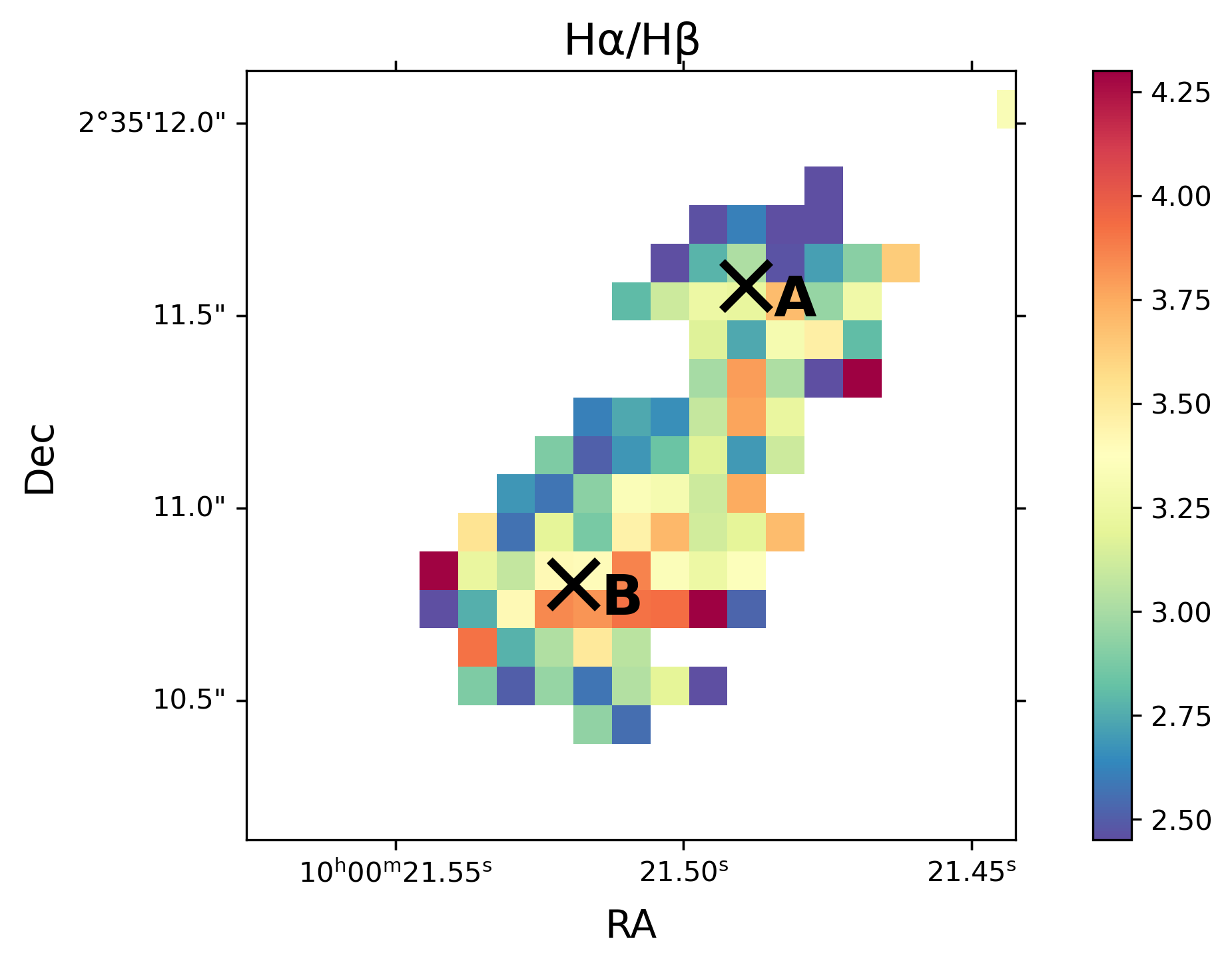}
\caption{\small 
H$\alpha$/H$\beta$ map derived from the JWST/NIRSpec data cube. The map is shown at the original spatial resolution, highlighting the intrinsic variations in the Balmer decrement across the system. Regions A and B, marked with black crosses, exhibit distinct line ratios. Compared to the MUSE convolved map (see Figure~\ref{lyaHa_fluxmap}, bottom panel), the native-resolution data reveal a sharper contrast between these regions, indicating that the smooth gradient observed after convolution is primarily driven by spatial mixing of flux. }
\label{hahb_notconvolved}
\end{figure}

\clearpage

\end{appendix}
\end{document}